\documentclass[a4paper,12pt]{article}
\usepackage{amsfonts}
\usepackage[usenames]{color}
\usepackage{graphicx}
\usepackage{amssymb}
\usepackage{amsthm}
\usepackage{amsmath}
\usepackage{listings}
\usepackage{multicol}
\usepackage{eurosym}
\usepackage{subcaption}
\usepackage{a4}
\usepackage{setspace}
\usepackage[outdir=./]{epstopdf}
\usepackage{listings}
\usepackage{float}
\usepackage{framed}
\usepackage[margin=1.8cm]{geometry}
\usepackage[latin1]{inputenc}
\usepackage[bottom]{footmisc}
\usepackage{rotating}
\usepackage{bbm}
\usepackage{titling}
\usepackage[colorlinks=true, urlcolor=black, linkcolor=blue,citecolor=blue]{hyperref}
\usepackage{natbib}
\usepackage[flushleft]{threeparttable}
\usepackage[labelfont=bf]{caption}
\usepackage[export]{adjustbox}
\usepackage{epstopdf}
\usepackage[affil-it]{authblk}

\definecolor{royalazure}{rgb}{0.0, 0.2, 0.65}
\definecolor{amaranth}{rgb}{0.9, 0.17, 0.31}
\definecolor{darkgreen}{RGB}{0.0, 128, 0.0}
\definecolor{darkgoldenrod}{rgb}{0.72, 0.53, 0.04}

\setcitestyle {aysep={,}} 

 \allowdisplaybreaks 
\theoremstyle {plain}     
\newtheorem{example}{Example}
\newtheorem{rem}{Remark}
\begin{document}

\title{\textbf{Seasonality in Mixed Causal-Noncausal Processes}}
\author[1]{Tom\'{a}s del Barrio Castro}
\author[2]{Alain Hecq} 
\author[3]{Sean Telg\thanks{%
Correspondence to: Sean Telg, Vrije Universiteit Amsterdam, Department of Econometrics and Data Science, De Boelelaan 1105, 1081 HV Amsterdam, The Netherlands. E-mail: j.m.a.telg@vu.nl. Tel: +31 20 59 89449.}}

\affil[1]{University of the Balearic Islands}
\affil[2]{Maastricht University}
\affil[3]{Vrije Universiteit Amsterdam}
%\affil[4]{Tinbergen Institute}

\maketitle

\begin{abstract}
\noindent This paper investigates the role of complex and negative roots in mixed causal-noncausal autoregressive (MAR) models. Using partial fraction decompositions, we show that seasonal roots can always be isolated in the moving average representation of purely causal and noncausal AR models. We find that this result extends to the MAR model, which means that no new joint seasonal effects can be generated despite the multiplicative structure of the causal and noncausal polynomials. This results has important consequences for the MAR model selection procedure and these are extensively studied in a Monte Carlo simulation study. An empirical application on COVID-19 and soybean data illustrates the main findings of the paper. \\
\newline 
\noindent \textbf{Keywords:} noncausality, seasonality, non-Gaussian errors. \newline
\noindent \textbf{JEL codes:} C22, E31, E37.

\let\thefootnote\relax\footnote{The authors gratefully acknowledge financial support from Grant PID2023-150095NB-C43, funded by MICIU/AEI/10.13039/501100011033 and, where applicable, by ERDF/EU.}

\end{abstract}

\newpage

\section{Motivation}\label{sec:Int}
The mixed causal-noncausal autoregressive (MAR) 
model is a time series model driven by non-Gaussian noise that contains both lag and lead polynomial components. It has received considerable attention in the literature over the past years, in particular because of its ability to capture nonlinear dynamics induced by locally explosive episodes and spikes (see e.g., \citealp{GourierouxandZakoian2017, HecqandVoisin2020}). The main focus of these modeling exercises has been on the identification of (positive) bubble phenomena in financial and macroeconomic time series. There are, however, many time series that exhibit nonlinear patterns that are vastly different from temporary sharp increases followed by a crash. For example, one can observe an increase in the volatility of stock returns that brutally stops. Alternatively, one could encounter gross domestic product or inflation with strong periodic behavior in the form of seasonal oscillations, especially in the raw series. To model such features we emphasize the  existence of negative and complex roots in the real-valued causal and noncausal polynomials, which we call seasonality. 

The presence of such roots is well-understood in conventional autoregressive models, but it is not obvious how they extend to MAR models. It is well-known that the same roots in the backward- and forward-looking polynomial may generate different dynamics \citep{GourierouxandJasiak2016}. However, due to the multiplicative structure of the causal and noncausal polynomials, it is not directly clear how seasonality propagates through the system. Moreover, roots may appear as pairs of complex conjugates, which has consequences for the estimation and selection of MAR models, which are generally non-nested and may suffer from the well-studied bimodality issue \citep{Hecqetal2016, Becetal2020}. In this paper, we extensively address these issues. 

The inclusion of a noncausal component with seasonal roots offers the possibility to generate a richer set of dynamics than conventional causal autoregressive models can. Using partial fraction decompositions, we find that all roots associated to seasonal frequencies can be isolated and uniquely assigned to either the backward- or forward-looking part of the MAR model's moving average representation. This means that no additional seasonal effects can be generated through the multiplicative structure of the causal and noncausal polynomials. The procedure of first determining the total autoregressive order using the pseudo-causal model representation\footnote{This model is also referred to as the weak form or the second-order equivalent (SOE) representation of the process in the literature (see e.g. \citealp{FriesandZakoian2019}).} therefore remains valid. The roots that are recovered from this model provide a good basis to either estimate the strong form\footnote{That is, a representation of the process that has $i.i.d.$ disturbances.} directly or as starting values of estimation techniques such as approximate maximum likelihood (AML). In the first case, estimation is rather straightforward as it boils down to matching the correct roots to the causal and noncausal polynomial. However, to determine the correct causal and noncausal orders, one requires a criterion such as extreme residuals clustering \citep{FriesandZakoian2019}. Model selection is easier in the second case, since one chooses the model that maximizes the value of the log-likelihood function at the estimated parameters. A disadvantage lies in the fact that a parametric assumption on the error distribution is required. 

In any case, we argue that the presence of seasonal roots may simplify the model selection approach. More specifically, roots that appear as pairs of complex conjugates in the pseudo-causal model need to be supplied jointly to either the causal or noncausal polynomial. This feature reduces the number of feasible options. For example, if the pseudo-causal model is an autoregressive process of order two, then the strong form is either a purely causal or purely noncausal model since a MAR(1,1) specification 
%with causal and noncausal order equal to one 
is no longer possible. Moreover, we advocate the use of the MAR model where roots may be seasonal instead of explicitly formulating a multiplicative seasonal model, since the latter can be shown to be a restricted version of the former. 

The paper is organized as follows. Section \ref{sec:ARmodel} introduces the notion of seasonal roots in pure and mixed autoregressive models and shows that factors associated to different frequencies can be isolated. In Section \ref{sec:EstModSel}, we study the consequences of these findings in terms of estimation and model selection. An extensive Monte Carlo simulation study in Section \ref{sec:SimStudy} confirms the theoretical findings. Section \ref{sec:EmpApp} consists of two empirical illustrations on COVID-19 data of Belgium and Italy and soybean prices. Section \ref{sec:Conclusion} concludes. The Appendix collects additional material.

\section{The Autoregressive Model with Seasonality}\label{sec:ARmodel}
In this section, we study how to identify seasonality within autoregressive models. We first consider the purely causal AR model, show that seasonal effects can be isolated 
using a partial fraction decomposition and argue that this procedure is fully symmetric for purely noncausal models. We proceed to show that similar results hold for MAR models,
which means that the causal and noncausal components cannot jointly create new seasonal effects. Lastly, we consider two different extensions to the MAR model.

\subsection{Purely Causal and Noncausal Models}
We start by focusing on purely causal autoregressive processes and continue to show that these results are also applicable to their purely noncausal counterparts. That is, we consider the
stationary AR process $\{y_{t}\}_{t \in \mathbb{Z}}$ of order $p \in \mathbb{N}$ observed during a general number of seasons per year, equal to $S$:\footnote{Note that we do not restrict the analysis to seasonal autoregressive process, denoted AR($p$) with $p \leq S$, 
as in \cite{DelBarrioCastroetal2019}, which considers the factorization used in seasonal
unit roots papers (see also \citeauthor{delBarrioCastroetal2012}, \citeyear{delBarrioCastroetal2012} and 
\citeauthor{Smithetal2009}, \citeyear{Smithetal2009}). See Appendix A for more details.} 
\begin{equation}
a^{\ast}(L)y_{t}=\varepsilon^{\ast}_{t}, \label{eq:ar}
\end{equation}%
with $a^{\ast}(z):= 1-\sum_{j=1}^{p}a_{j}^{\ast }z^{j}$ having all roots strictly outside the unit circle and $L$ representing the lag operator such that $L^{k}y_{t}=y_{t-k}$ and $\{\varepsilon^{\ast}_{t}\}_{t \in \mathbb{Z}}$ is an $i.i.d.$ sequence. Note that the results in this section about purely causal and noncausal model are also valid for the less stringent assumption that $\{\varepsilon^{\ast}_{t} \}_{t \in \mathbb{Z}}$ is a white noise sequence. The choice for an $i.i.d.$ sequence is solely made because we want to study seasonality within the framework of mixed causal-noncausal models afterwards. 

\subsubsection{Seasonal Roots}\label{sec:seasonalroots}
We can factorize the polynomial $a^{\ast}(z)$ in monomials associated to the
inverse roots $\alpha_{k}$, $k=1,2,\ldots, p,$ of $a^{\ast}(z)$, which could be real or complex valued:
\begin{equation}
a^{\ast}(z)=\prod_{k=1}^{p}\left( 1-\alpha _{k}z\right).  \label{eq:polynom}
\end{equation}% 
Given that $a^{\ast}(z)$ is a real valued
coefficient polynomial, the complex valued inverse roots appear in pairs of
complex conjugates. Define the inverse root $\alpha_{k} :=\alpha _{k}^{R}+{\mathrm{i}}\alpha _{k}^{I}$, 
where $\alpha _{k}^{R}$ is the real part of $\alpha _{k}$ ($\mbox{Re}\left[ \alpha _{k}\right]
:=\alpha _{k}^{R}$), $\alpha _{k}^{I}$ is the imaginary part of $\alpha _{k}$ ($%
\mbox{Im}\left[ \alpha _{k}\right] :=\alpha _{k}^{I}$) and ${\mathrm{i}} :=%
\sqrt{-1}$. To allow for possible seasonal behavior, we focus on
the exponential form to represent complex valued inverse roots, i.e. $%
\alpha _{k}=\rho _{k}e^{{\mathrm{i}}\omega _{k}}$, where $\rho _{k}$ is the
modulus of the complex number defined as $\rho _{k} := \sqrt{\left( \alpha _{k}^{R}\right)
^{2}+\left( \alpha _{k}^{I}\right) ^{2}}$ and $\omega _{k}$ is the argument
of the complex number, i.e. $\omega _{k} := \arctan \left( \alpha _{k}^{I}/\alpha
_{k}^{R}\right)$.

\begin{rem}
Note that the argument $\omega _{k}=\arctan \left( \alpha _{k}^{I}/\alpha
_{k}^{R}\right) $ represents the angle in radians that $\alpha _{k}=\alpha
_{k}^{R}+\mathrm{i}\alpha _{k}^{I}$ makes with the positive real axis when $\alpha
_{k}$ is interpreted as a vector bound from the origin. It is not
uniquely defined since the tangent of $\alpha _{k}^{I}/\alpha _{k}^{R}$ is
not affected when integer multiples of $2\pi$ are added to or subtracted from $%
\alpha _{k}^{I}/\alpha _{k}^{R}$. Therefore, it is better to use the principal
argument, i.e. the angle in the interval $\left( -\pi ,\pi \right] $ which
satisfies the definition. It can be obtained using the``2-argument arctangent" 
function $\arctan2(\alpha _{k}^{R},\alpha _{k}^{I})$, which we will characterize
later.
\end{rem}

The polynomial $a^{\ast}(z)$ is composed of three types of roots. Real valued inverse roots in 
$\left( 1-\rho_{k}e^{{\mathrm{i}}\omega _{k}}z\right)$ appear when $%
\omega _{k}=0$ or $\omega_{k}=\pi$, which yield $\left( 1-\rho_{k}z\right) $ and 
%$\left( 1-\rho _{k}\left( -1\right) L\right) =%
$\left(1+\rho_{k}z\right)$ respectively.\footnote{This follows directly as $\rho_{k}e^{\pm {\mathrm{i}}0} = \rho _{k}$ and
$\rho_{k}e^{\pm{\mathrm{i}}\pi}=-\rho_{k}$.} 
%In particular $\omega _{k}=0$, 
The factor $\left( 1-\alpha _{k}z\right) $ is associated to the zero frequency,
and the factor $\left( 1+\alpha _{k}z\right) $ is associated to the Nyquist
frequency $\pi$, i.e. oscillations that complete a full cycle every two
periods. If we have complex inverse roots in $a^{\ast}(z)$, they will appear in complex conjugate pairs 
$\left( 1-\alpha_{k}z\right) \left( 1-\bar{\alpha}_{k}z\right)
=\left( 1-\rho_{k}e^{{\mathrm{i}}\omega _{k}}z\right) \left( 1-\rho_{k}e^{-%
{\mathrm{i}}\omega_{k}}z\right) =\left( 1-2\cos \left( \omega_{k}\right)
\rho _{k}z+\rho_{k}^{2}z^{2}\right) $. Note that the term $\left( 1-\rho _{k}e^{\pm {%
\mathrm{i}}\omega_{k}}z\right) $ is associated to frequency $\omega
_{k}$, which are oscillations that complete a full cycle every $2\pi /\omega
_{k}$ periods.  Hence, we can account for both seasonal and
cyclical behavior in $a^{\ast}(z)$ using an alternative representation to \eqref{eq:polynom}: 
\begin{equation}\label{eq:polynom2}
a^{\ast}(z) =\prod_{k=1}^{p}\left( 1-\rho _{k}e^{{\mathrm{i}}\omega _{k}}z\right),
\end{equation}%
in which it is understood that
complex valued roots for $\omega _{k} \notin \{ 0,\pi \}$ appear as
a pair of complex conjugates $\left( 1-\rho _{k}e^{{\mathrm{i}}\omega
_{k}}z\right) \left( 1-\rho _{k}e^{-{\mathrm{i}}\omega _{k}}z\right) $.
Thus, seasonal behavior happens for real valued inverse roots associated to
factor $\left( 1+\rho _{k}z\right) $ and two pairs of complex conjugates $%
\left( 1-\rho _{k}e^{\pm {\mathrm{i}}\omega _{k}}z\right) $ with $\omega
_{k} \in \{ 0,\pi \} $ and $\omega _{k}=2\pi k/S$ with $k=1,2,\ldots
,\left\lfloor \left( S-1\right) /2\right\rfloor $, with $\left\lfloor
.\right\rfloor $ denoting the integer part of its argument. Since we do not restrict our attention to seasonal AR($p$) models such that $p \leq S$, we can have multiple roots at both the zero and seasonal frequencies.

\subsubsection{Combinations of Roots}\label{sec:comroots}
Similar to \cite{DelBarrioCastroetal2019}, we use the partial fraction decomposition of the polynomial associated to an autoregressive process to investigate the presence of different combinations of roots. Applying results from \citeauthor{Pollock1999} 
(\citeyear{Pollock1999}, Chapter 3) to \eqref{eq:polynom2}, we obtain
\begin{equation} \label{eq:RA1}
\frac{1}{\left( 1-\rho _{k}e^{\pm {\mathrm{i}}\omega _{k}}L\right) \left(
1-\rho _{j}e^{\pm {\mathrm{i}}\omega _{j}}L\right) }=\frac{\rho _{k}e^{\pm {%
\mathrm{i}}\omega _{k}}/\left( \rho _{k}e^{\pm {\mathrm{i}}\omega _{k}}-\rho
_{j}e^{\pm {\mathrm{i}}\omega _{j}}\right) }{\left( 1-\rho _{k}e^{\pm {%
\mathrm{i}}\omega _{k}}L\right) }+\frac{\rho _{j}e^{\pm {\mathrm{i}}\omega
_{j}}/\left( \rho _{j}e^{\pm {\mathrm{i}}\omega _{j}}-\rho _{k}e^{\pm {%
\mathrm{i}}\omega _{k}}\right) }{\left( 1-\rho _{j}e^{\pm {\mathrm{i}}\omega
_{j}}L\right)},  
\end{equation}
and note that this general case \eqref{eq:RA1} covers all
possible combinations of roots. That is, 
\begin{subequations}
\begin{align}
\frac{1}{\left( 1-\rho _{k}L\right) \left( 1+\rho _{j}L\right) } &= \frac{%
\rho _{k}/\left( \rho _{k}+\rho _{j}\right) }{\left( 1-\rho _{k}L\right) }+%
\frac{\rho _{j}/\left( \rho _{j}+\rho _{k}\right) }{\left( 1+\rho
_{j}L\right) },  \label{eq:RA2} \\
\frac{1}{\left( 1-\rho _{k}L\right) \left( 1-\rho _{j}e^{\pm {\mathrm{i}}%
\omega _{j}}L\right) } &= \frac{\rho _{k}/\left( \rho _{k}-\rho _{j}e^{\pm {%
\mathrm{i}}\omega _{j}}\right) }{\left( 1-\rho _{k}L\right) }+\frac{\rho
_{j}e^{\pm {\mathrm{i}}\omega _{j}}/\left( \rho _{j}e^{\pm {\mathrm{i}}%
\omega _{j}}-\rho _{k}\right) }{\left( 1-\rho _{j}e^{\pm {\mathrm{i}}\omega
_{j}}L\right) },  \label{eq:RA3} \\
\frac{1}{\left( 1+\rho _{k}L\right) \left( 1-\rho _{j}e^{\pm {\mathrm{i}}%
\omega _{j}}L\right) } &= \frac{\rho _{k}/\left( \rho _{k}+\rho _{j}e^{\pm {%
\mathrm{i}}\omega _{j}}\right) }{\left( 1+\rho _{k}L\right) }+\frac{\rho
_{j}e^{\pm {\mathrm{i}}\omega _{j}}/\left( \rho _{j}e^{\pm {\mathrm{i}}%
\omega _{j}}+\rho _{k}\right) }{\left( 1-\rho _{j}e^{\pm {\mathrm{i}}\omega
_{j}}L\right) },  \label{eq:RA4} \\
\frac{1}{\left( 1-\rho _{k}e^{-{\mathrm{i}}\omega _{k}}L\right) \left(
1-\rho _{k}e^{{\mathrm{i}}\omega _{k}}L\right) } &= \frac{e^{-{\mathrm{i}}%
\omega _{k}}/\left( e^{-{\mathrm{i}}\omega _{k}}-e^{{\mathrm{i}}\omega
_{k}}\right) }{\left( 1-\rho _{k}e^{-{\mathrm{i}}\omega _{k}}L\right) }+%
\frac{e^{{\mathrm{i}}\omega _{k}}/\left( e^{{\mathrm{i}}\omega _{k}}-e^{-{%
\mathrm{i}}\omega _{k}}\right) }{\left( 1-\rho _{k}e^{{\mathrm{i}}\omega
_{k}}L\right)},  \label{eq:RA5}
\end{align}
\end{subequations}
where \eqref{eq:RA2} considers the real valued cases $\omega_{k} =0$ and 
$\omega_{j} = \pi$, \eqref{eq:RA3}-\eqref{eq:RA4} a mixture of
real and complex valued roots and \eqref{eq:RA5} a complex conjugate pair.
These results imply that it is possible to express the process \eqref {eq:ar}
in terms of a partial fraction decomposition by writing it
in its moving average (MA) representation, i.e. $y_{t}=a(L)^{-1}\varepsilon^{\ast}_{t}$, 
and concluding that it can always be represented in the following way: 
\begin{equation}
y_{t}=\left( \sum_{k=1}^{p}\left[ \frac{d_{k}}{\left( 1-\rho _{k}e^{{\mathrm{%
i}}\omega _{k}}L\right) }\right] \right) \varepsilon^{\ast}_{t}.
\label{eq:parfracdecomp}
\end{equation}%
Note that in (\ref{eq:parfracdecomp}) the terms $d_{k}/\left( 1-\rho _{k}e^{{%
\mathrm{i}}\omega _{k}}L\right)$ appear in pairs of complex conjugate terms for $\omega _{k} \notin \{0,\pi\}$ as in expression \eqref{eq:RA5}. Note that based on \eqref{eq:RA1}, which covers the cases \eqref{eq:RA2}-\eqref{eq:RA5}, it is possible to compute the value
of the terms $d_{k}$ in \eqref{eq:parfracdecomp}. 
%depending on the factors present in \eqref{eq:polynom2}. 
\citeauthor{Pollock1999} (\citeyear{Pollock1999}, Chapter 3) provides a simple and quick method
to obtain the coefficients $d_{k}$ of the partial fraction decomposition
of $a(L)^{-1}$. This extends to the case of
inverse roots with multiplicity of at least two, which we characterize in
Example \ref{ex:pollockmethod}.

\begin{example}\label{ex:pollockmethod}
We can show that with multiplicity equal to two of one of the inverse roots, we have
\begin{equation*}
\frac{1}{\left( 1-\alpha L\right) ^{2}\left( 1-\beta L\right) }=\frac{d_{1}}{%
\left( 1-\alpha L\right) ^{2}}+\frac{d_{2}}{\left( 1-\alpha L\right) }+\frac{%
d_{3}}{\left( 1-\beta L\right)},
\end{equation*}%
and continue to compute $d_{1}$, $d_{2}$ and $d_{3}$. The first step is
to write the left-hand side in its partial fraction decomposition:
\begin{equation}\label{eq:partfracexample1}
\frac{1}{\left( 1-\alpha z\right) ^{2}\left( 1-\beta z\right)} = \frac{\eta_{1}(z)}{\left( 1-\alpha z\right) ^{2}} + \frac{\eta_{2}(z)}{\left( 1-\beta z\right)}.
\end{equation}
Let us focus on the first fraction on the right-hand side. Note that we can write $\eta_{1}(z) = d_{1} + d_{2}(1-\alpha z)$ following Pollock (1999, p. 61). Substituting this expression into \eqref{eq:partfracexample1} and multiplying the complete expression by $(1-\alpha z)^{2}$ yields
\begin{equation}\label{eq:partfracexample2}
\frac{1}{\left( 1-\beta z\right)} = d_{1} + d_{2}\left( 1-\alpha z\right) + (1-\alpha z)^{2} \frac{\eta_{2}(z)}{\left( 1-\beta z\right)},
\end{equation}
which yields $d_{1} = 1/(1-\beta / \alpha)$ when setting $z = 1/\alpha$. Subsequently, we take the derivative of \eqref{eq:partfracexample2} with respect to $z$ to find $d_{2}$:
\begin{equation*}
\frac{\beta}{(1-\beta z)^{2}} = -\alpha d_{2} + \frac{\partial}{\partial z} \left[ (1-\alpha z)^{2} \frac{\eta_{2}(z)}{\left( 1-\beta z\right)} \right].
\end{equation*}
By setting $z = 1/ \alpha$ we note that the last expression equals zero and find $d_{2} = \beta / [-\alpha(1-\beta / \alpha)^{2}]$. Analogously, we can treat the second fraction on the right-hand side of \eqref{eq:partfracexample1} and obtain $d_{3} = 1/(1-\alpha / \beta)^{2}$.
\end{example}

\begin{figure}[tbp]
\captionsetup[subfigure]{labelformat=empty}
\centering
\begin{adjustbox}{minipage=\linewidth,scale=0.95}
\begin{subfigure}[b]{\textwidth}
								\centering
                \includegraphics[width=\textwidth]{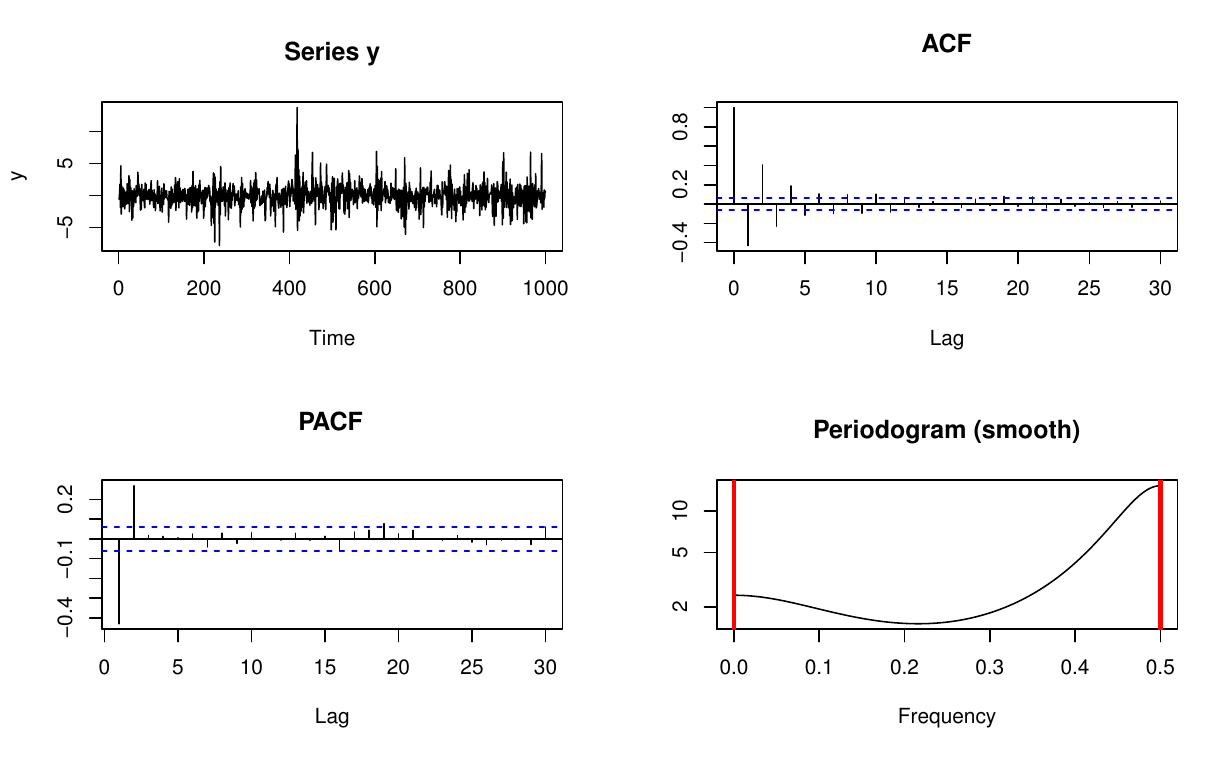}
                \caption{(a) Process with real-valued roots at the zero and Nyquist frequency}
                \label{fig:ZeroNyquistAR}
\end{subfigure}
\vskip\baselineskip
\begin{subfigure}[b]{\textwidth}
								\centering
                \includegraphics[width=\textwidth]{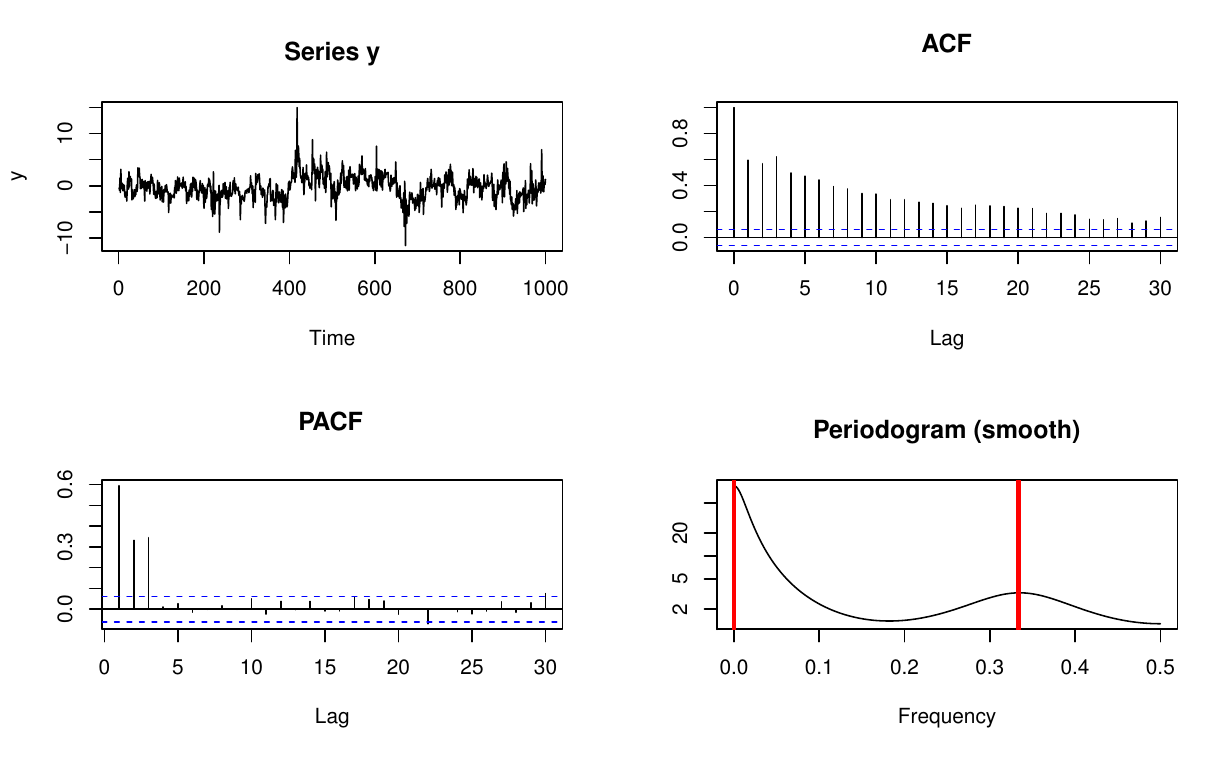}
                \caption{(b) Process with one real-valued root at the zero frequency and complex-valued roots at frequency $\frac{2}{3}\pi$}
                \label{fig:ZeroComplexAR}
\end{subfigure} 
\end{adjustbox}
\caption{Simulated AR processes including ACF, PACF and smoothed periodogram}
\label{fig:dataperiodAR}
\begin{minipage}{\textwidth}\footnotesize
    Note: Frequencies have been mapped from $\left[ 0,\pi \right]$ to $\left[0,\frac{1}{2} \right]$, as the periodograms are displayed in the latter interval. The process in (a)
    represents an AR(2) process, with a combination of one root at the zero frequency and the other root at the Nyquist frequency $\pi$. Process (b) is an AR(3) process that combines a root at the zero frequency and roots that appear in a pair of complex conjugates.  It considers the case $S=6$, $k= \left\lfloor \left( 6-1\right) /2\right\rfloor = 2$ leading to frequency $\frac{2}{3}\pi$. The red lines in the periodograms highlight the expected peaks based on the chosen specifications.
\end{minipage}
\end{figure}

In conclusion, from \eqref {eq:parfracdecomp} it follows that the
factors that could cause power in the spectrum at seasonal
frequencies is restricted to two terms: $(i)$ $1/(1+\rho _{k}z)$ associated
with the Nyquist frequency $\pi$ and $(ii)$ $1/(1-\rho _{k}e^{\pm {\mathrm{i}}%
\omega _{k}}z)$ associated with the harmonic frequencies $\omega _{k}$ and $%
2\pi -\omega _{k}$. We illustrate this in Figure \ref{fig:dataperiodAR}, which displays two simulated autoregressive processes of length $T=1000$, their corresponding autocorrelation function (ACF), partial autocorrelation function (PACF), and smoothed periodogram with red vertical lines indicating the frequencies corresponding to the roots of the AR polynomial, mapped from the interval $[0, \pi]$ to $[0,\frac{1}{2}]$. The error term $\{\varepsilon^{*}_{t} \}_{t=1}^{T}$ of the AR processes follows a non-standardized Student's $t$ distribution, denoted $t(\nu,\sigma)$, with degrees of freedom $\nu = 3$ and scale parameter $\sigma = 1$, and different configurations of roots are chosen in each case. In Figure \ref{fig:ZeroNyquistAR}, we consider an AR(2) process with inverse roots $\alpha_{1} = 0.4$ and $\alpha_{2} = -0.7$, yielding $a^{*}(z) = 1 + 0.3z - 0.28z^{2}$. Since one root of this polynomial is at the zero frequency and the other at the Nyquist frequency, we expect the spectrum to peak at the start and the end, which is indeed the case. Both the ACF and PACF show an oscillating effect, which reveals the presence of the seasonal root. As only the first two lags are significantly different from zero at a $5\%$ significance level in the PACF, we are thus able to reveal the main structure of the process using these measures combined. Figure \ref{fig:ZeroComplexAR} represents an AR(3) process with one root at the zero frequency and the other roots appearing as a pair of complex conjugates. More specifically, we consider $a^{\ast}(z) = 1 - 0.3z - 0.18z^{2} - 0.324z^{3}$ which corresponds to inverse roots $\alpha_{1} = 0.9$ and $\alpha_{2,3} = (-0.833 \pm 1.443 \mathrm{i})^{-1}$ belonging to the frequencies zero and $\frac{2}{3}\pi$ respectively. Once again, we see that the spectrum peaks at the expected frequencies. The wave-form in the ACF tacitly reveals the presence of seasonal roots, while the PACF has three significant lags and therefore correctly identifies the autoregressive order. It is interesting to notice that the presence of seasonal roots is often not directly visible from the time series trajectories. Overall, they can look identical to regular AR processes with different degrees of persistency. This emphasizes the need for tools that can detect different types of roots.

\subsubsection{Purely Noncausal Models}\label{sec:purenc}
If we replace the lag operator in \eqref{eq:ar} by a lead operator, we obtain a purely noncausal process
\begin{equation*} 
a(L^{-1})y_{t}=\varepsilon^{*}_{t},
\end{equation*} 
where $L^{-k}y_{t}=y_{t+k}$ and the corresponding polynomial $a(z) := 1-\sum_{j=1}^{p}a_{j}^{\ast }z^{j}$
still has all roots strictly outside the unit circle.
By exact symmetry of the model, i.e., the process only differs in terms of the used operator, we note that the derived findings in Section \ref{sec:comroots} are fully analogous. This means that also noncausal processes can be represented as in \eqref{eq:parfracdecomp} when we replace $L$ by $L^{-1}$. The main reason to study these processes lies in their ability to mimic certain non-linear features in data that causal counterparts cannot. Existing literature typically compares the processes based on roots at the zero frequency, which encompasses the often-studied case of speculative bubbles.

\begin{figure}[ht]
\captionsetup[subfigure]{labelformat=empty}
\centering
\begin{adjustbox}{minipage=\linewidth,scale=0.9}
\begin{subfigure}[b]{\textwidth}
								\centering
                \includegraphics[width=\textwidth]{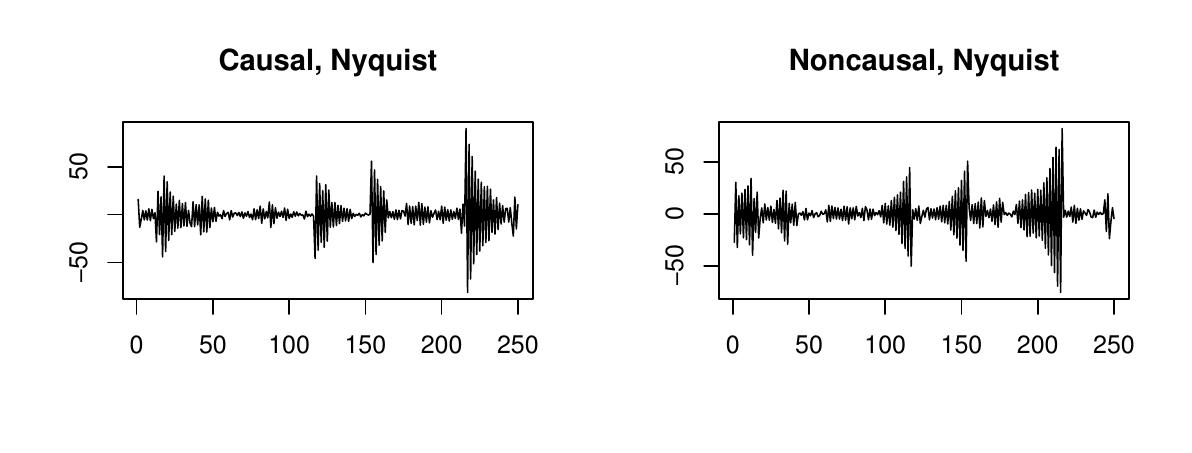}
                \caption{(a) Processes with root at Nyquist frequency $\rho_{k} = -0.9$}
                \label{fig:PureNyquist}
\end{subfigure}
\vskip\baselineskip
\begin{subfigure}[b]{\textwidth}
								\centering
                \includegraphics[width=\textwidth]{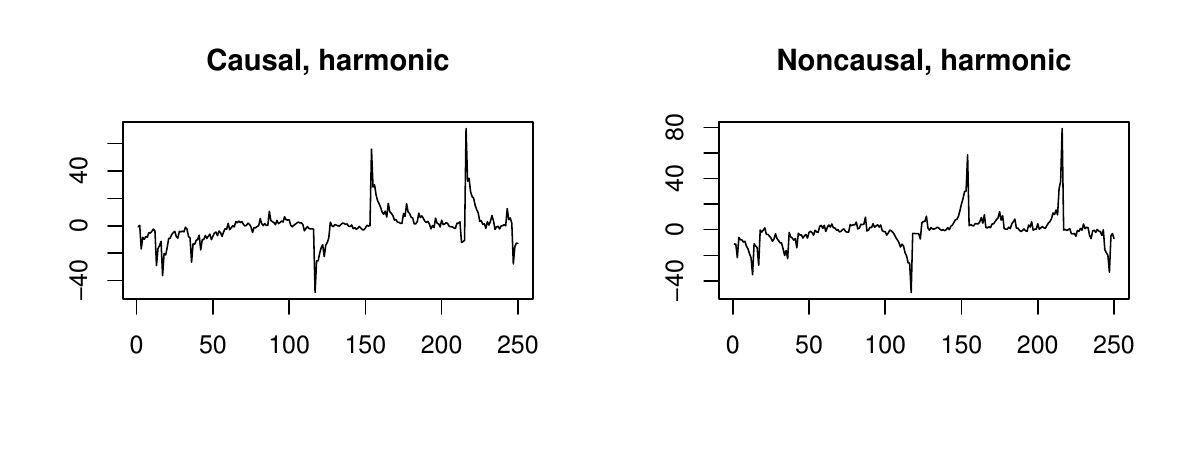}
                \caption{(b) Processes with complex-valued roots at frequency $\frac{2}{3}\pi$ and $\rho_{k} = 0.5$}
                \label{fig:PureHarmonic}
\end{subfigure} 
\end{adjustbox}
\caption{Simulated causal (left) and noncausal (right) AR processes with seasonal roots}
\label{fig:PureAR}
\end{figure}

In Figure \ref{fig:PureAR} we show trajectories of causal and noncausal processes for the case of seasonal roots. The error term is assumed to follow a standard Cauchy error distribution, which is often used to generate locally explosive dynamics. Figure \ref{fig:PureNyquist} considers an AR(1) with an inverse root at the Nyquist frequency $\pi$, i.e., $\alpha_{k} = -0.9$. We observe the typical oscillating effect with the main difference that the extreme shock fades out for the causal case (left), while it gradually amplifies for the noncausal case (right). The latter case could be interpreted as a seasonal bubble, in the sense that there is temporary explosive behavior followed by a return to the baseline path. The bubbles resemble periods of short-term increases in volatility similar to conditional heteroskedasticity, while speculative bubbles generated by roots at the zero frequency only seem to affect the level of the series. Figure \ref{fig:PureHarmonic} shows that complex-valued inverse roots $\alpha_{k} = \rho_{k} e^{\mathrm{i}\omega_{k}}$ with $\rho_{k} = 0.5$ at the harmonic frequencies $\omega _{k} = \frac{2}{3}\pi$ (and $2\pi - \omega_{k}$) are also able to generate causal and noncausal trajectories that are almost symmetric. However, the noncausal case reveals that bubbles can be generated which resemble the ones that are due to roots at the zero frequency. Thus, seasonal behavior is not always explicit from the trajectory. Depending on the choice of error distribution and parameter values, causal and noncausal dynamics might also be more difficult to disentangle. Interestingly, the causal and noncausal processes in these figures are fully identical in terms of second-order properties. This means that we cannot distinguish them based on the ACF, PACF or the spectrum. However, their ability to generate different types of dynamics makes a convincing case for combining causal and noncausal behavior in autoregressive processes.

\subsection{Mixed Causal-Noncausal Models}\label{sec:MAR}
Up until now, we have only considered autoregressive processes that 
have a one-sided MA($\infty$) representation. That is, since $a^{*}(z)$ in \eqref{eq:ar} 
has all roots outside the unit circle, the strictly stationary solution of $\{y_{t}\}_{t \in \mathbb{Z}}$ takes the form of a one-sided moving average given by $y_{t} = \sum_{j=0}^{\infty} \zeta_{j} \varepsilon^{*}_{t-j}$. 
% Note: Theorem 3.1.1. in Brockwell and Davis (1991) shows that the summability condition $\sum_{j=0}^{\infty} |\zeta_{j}| < \infty$ is
% automatically fulfilled when all roots are outside the unit circle.
As alluded to in Section \ref{sec:purenc}, richer dynamic patterns can be modeled if the causality assumption
is abandoned and $a^{*}(z)$ is allowed to have roots both inside
and outside the unit circle.\footnote{The only case we exclude is the
presence of unit roots: $a^{*}(z) = 0$ for $|z| = 1$.} Therefore, we continue to study the mixed causal-noncausal process which admits a two-sided MA($\infty$) representation
\begin{equation}\label{eq:marep}
y_{t} = \sum_{j=-\infty}^{\infty} \xi_{j} \epsilon_{t+j},
\end{equation}  
where \cite{BrockwellandDavis1991} detail the appropriate summability
conditions on the sequence $\{ \xi_{j} \}_{j \in \mathbb{Z}}$ in both the finite and
infinite variance framework for the errors.

\subsubsection{Model Representation}
The mixed causal-noncausal model has two different representations in the literature.
\cite{Breidtetal1991} consider a process $\{ y_{t} \}_{t \in \mathbb{Z}}$
\begin{equation}\label{eq:mar.alt}
a(L)y_{t} = \epsilon_{t},
\end{equation}
where $a(z)$ is a polynomial of order $p=r+q$, which has $r$ roots outside 
and $q$ roots inside the unit circle. Since we have $a(z) \neq 0$ for $|z| = 1$, we can write $a(z) = \phi(z)\varphi^{\ast}(z)$, where $\phi(z) := 1 - \sum_{j=1}^{r} \phi_{j}z^{j}$ and $\varphi^{\ast}(z) := 1 -
\sum_{j=1}^{q} \varphi^{\ast}_{j}z^{j}$ collect the well-behaved and ill-located roots, respectively. We can express
$\varphi^{\ast}(z)$ in terms of the polynomial $\varphi(z^{-1})$, whose roots are the reciprocals of those of $\varphi^{\ast}(z)$
and are therefore located strictly outside the unit circle:
\begin{equation*}
\varphi^{\ast}(z) = - \varphi^{\ast}_{q} z^{q}\varphi(z^{-1}),
\end{equation*}
with $\varphi^{*}_{q-j}/\varphi^{*}_{q} = -\varphi_{j}$ for $j=1,...,q-1$ and $1/\varphi^{*}_{q} = \varphi_{q}$
(and thus $\varphi_{q} \neq 0$). Hence, if we define $\varepsilon_{t} = (-1/\varphi^{*}_{q})\epsilon_{t+q}$
, which is still $i.i.d.$ as it is simply a rescaled and time-shifted version of $\{\epsilon_{t}\}_{t \in \mathbb{Z}}$,
we obtain
\begin{equation}  \label{eq:mar}
\phi (L)\varphi (L^{-1})y_{t}=\varepsilon _{t},
\end{equation}
where both polynomials have their zeros outside
the unit circle such that
$\phi(z) \neq 0$ for $|z| \leq 1$ and $\varphi(z) \neq 0$ for $|z| \leq 1$.
This is the well-known mixed causal-noncausal autoregressive (MAR) model
as introduced by \cite{LanneandSaikkonen2011}. We denote the model as MAR($r,q$), where the first entry represents the causal order $r \in \mathbb{N}$ and the second entry the noncausal order $q \in \mathbb{N}$. For identification purposes, $\{\varepsilon _{t}\}_{t \in \mathbb{Z}}$ is assumed to be a non-Gaussian $i.i.d.$ sequence.

Whereas both representations \eqref{eq:mar.alt} and \eqref{eq:mar} are 
equally valid in the univariate framework, we study the MAR in multiplicative form estimated by AML in this paper.
The results can easily be rewritten into the other representation. An alternative semi-parametric approach for \eqref{eq:mar.alt} that is free of distributional assumptions would be the Generalized Covariance (GCov) framework proposed by \cite{GourierouxandJasiak2017, gourieroux2023generalized}.

\subsubsection{Combinations of Roots}
We proceed to show that the class of MAR($r,q$) models in \eqref{eq:mar} also admits a partial fraction
representation which allows for isolating seasonal components. We first note that the
extension of \eqref{eq:polynom}-\eqref{eq:polynom2} to the MAR case is given by 
\begin{equation*}
\phi (z)\varphi (z^{-1})=\left( \prod_{k=1}^{r}(1-\rho _{k}e^{{\mathrm{i}}%
\omega _{k}}z)\right) \left( \prod_{\ell =1}^{q}(1-\tilde{\rho}_{\ell }e^{{%
\mathrm{i}}\tilde{\omega}_{\ell }}z^{-1})\right),
\end{equation*}%
where $\tilde{\rho}_{\ell }$ and $\tilde{\omega}_{\ell }$ have the same
interpretation as $\rho_{k}$ and $\omega_{k}$. The tildes solely emphasize that
the terms are part of the noncausal polynomial. For illustrative purposes, we focus on MAR models where the
causal and noncausal components are combinations of factors at different
frequencies. Following \cite{GourierouxandJasiak2016}, 
it is possible to write
\begin{equation} \label{eq:GouJak}
\frac{1}{(1-\rho _{k}e^{{\mathrm{i}}\omega _{k}}L)(1-\tilde{\rho}_{\ell }e^{{%
\mathrm{i}}\tilde{\omega}_{\ell }}L^{-1})} 
%&= \frac{L}{(1-\rho _{k}e^{{%
%\mathrm{i}}\omega _{k}}L)(L-\tilde{\rho}_{\ell }e^{{\mathrm{i}}\tilde{\omega}%
%_{\ell }})}  \\ 
= L\left[ \frac{1}{(1-\rho _{k}e^{{\mathrm{i}}\omega _{k}}L)(L-\tilde{\rho}%
_{\ell }e^{{\mathrm{i}}\tilde{\omega}_{\ell }})}\right],
\end{equation}%
and with the term %$1/\left[ (1-\rho _{k}e^{{\mathrm{i}}\omega _{k}}L)(L-\tilde{\rho}%_{\ell }e^{{\mathrm{i}}\tilde{\omega}_{\ell }})\right]$ 
between the large square brackets we can
proceed as in \eqref{eq:RA1} to obtain a partial fraction decomposition given by%
\begin{equation}
\frac{1}{(1-\rho _{k}e^{{\mathrm{i}}\omega _{k}}L)(L-\tilde{\rho}_{\ell }e^{{%
\mathrm{i}}\tilde{\omega}_{\ell }})}=\frac{\rho _{k}e^{{\mathrm{i}}%
\omega _{k}}/\left( 1-\rho _{k}e^{\mathrm{i}\omega_{k}}\tilde{\rho}_{\ell
}e^{\mathrm{i}\tilde{\omega}_{\ell}}\right)}{(1-\rho _{k}e^{{\mathrm{i%
}}\omega _{k}}L)}+\frac{1 / \left( 1-\rho_{k}e^{\mathrm{i}\omega_{k}}
\tilde{\rho}_{\ell }e^{\mathrm{i}\tilde{\omega}_{\ell}}\right)}{(L-%
\tilde{\rho}_{\ell }e^{{\mathrm{i}}\tilde{\omega}_{\ell }})},
\label{eq:GouJak2}
\end{equation}%
which can be rewritten in a more familiar form that includes the lead operator 
by combining \eqref{eq:GouJak} and \eqref{eq:GouJak2}:
\begin{equation}
\frac{1}{(1-\rho _{k}e^{{\mathrm{i}}\omega _{k}}L)(1-\tilde{\rho}_{\ell }e^{{%
\mathrm{i}}\tilde{\omega}_{\ell }}L^{-1})}=\frac{1}{\left( 1-\rho _{k}e^{{%
\mathrm{i}}\omega _{k}}\tilde{\rho}_{\ell }e^{{\mathrm{i}}\tilde{\omega}%
_{\ell }}\right) }\left[ \frac{\rho _{k}e^{{\mathrm{i}}\omega _{k}}L}{%
(1-\rho _{k}e^{{\mathrm{i}}\omega _{k}}L)}+\frac{1}{(1-\tilde{\rho}_{\ell
}e^{{\mathrm{i}}\tilde{\omega}_{\ell }}L^{-1})}\right] .  \label{eq:RRA1}
\end{equation}%
Similar to the case of purely causal and noncausal models, this result
allows one to derive various combinations of roots. In particular, 
it is possible to obtain from \eqref{eq:RRA1} the following cases:%
\begin{subequations}
\begin{align}
\frac{1}{(1+\rho _{k}L)(1-\tilde{\rho}_{\ell }L^{-1})} &= \frac{1}{\left(
1+\rho _{k}\tilde{\rho}_{\ell }\right) }\left[ \frac{-\rho _{k}L}{(1+\rho
_{k}L)}+\frac{1}{(1-\tilde{\rho}_{\ell }L^{-1})}\right],  \label{eq:RRA4} \\
\frac{1}{(1-\rho _{k}L)(1+\tilde{\rho}_{\ell }L^{-1})} &= \frac{1}{\left(
1+\rho _{k}\tilde{\rho}_{\ell }\right) }\left[ \frac{\rho_{k}L}{(1-\rho
_{k}L)}+\frac{1}{(1+\tilde{\rho}_{\ell }L^{-1})}\right],  \label{eq:RRA7} \\
\frac{1}{(1-\rho _{k}L)(1-\tilde{\rho}_{\ell }e^{{\mathrm{i}}\tilde{\omega}%
_{\ell }}L^{-1})} &= \frac{1}{\left( 1-\rho _{k}\tilde{\rho}_{\ell }e^{{%
\mathrm{i}}\tilde{\omega}_{\ell }}\right) }\left[ \frac{\rho _{k}L}{(1-\rho
_{k}L)}+\frac{1}{(1-\tilde{\rho}_{\ell }e^{{\mathrm{i}}\tilde{\omega}_{\ell
}}L^{-1})}\right],  \label{eq:RRA2} \\
\frac{1}{(1+\rho _{k}L)(1-\tilde{\rho}_{\ell }e^{{\mathrm{i}}\tilde{\omega}%
_{\ell }}L^{-1})} &= \frac{1}{\left( 1+\rho _{k}\tilde{\rho}_{\ell }e^{{%
\mathrm{i}}\tilde{\omega}_{\ell }}\right) }\left[ \frac{-\rho _{k}L}{(1+\rho
_{k}L)}+\frac{1}{(1-\tilde{\rho}_{\ell }e^{{\mathrm{i}}\tilde{\omega}_{\ell
}}L^{-1})}\right],  \label{eq:RRA3} \\
\frac{1}{(1-\rho _{k}e^{{\mathrm{i}}\omega _{k}}L)(1-\tilde{\rho}_{\ell
}L^{-1})} &= \frac{1}{\left( 1-\rho _{k}e^{{\mathrm{i}}\omega _{k}}\tilde{%
\rho}_{\ell }\right) }\left[ \frac{\rho _{k}e^{{\mathrm{i}}\omega _{k}}L}{%
(1-\rho _{k}e^{{\mathrm{i}}\omega _{k}}L)}+\frac{1}{(1-\tilde{\rho}_{\ell
}L^{-1})}\right], \label{eq:RRA5} \\
\frac{1}{(1-\rho _{k}e^{{\mathrm{i}}\omega _{k}}L)(1+\tilde{\rho}_{\ell
}L^{-1})} &= \frac{1}{\left( 1+\rho _{k}e^{{\mathrm{i}}\omega _{k}}\tilde{%
\rho}_{\ell }\right) }\left[ \frac{\rho _{k}e^{{\mathrm{i}}\omega _{k}}L}{%
(1-\rho _{k}e^{{\mathrm{i}}\omega _{k}}L)}+\frac{1}{(1+\tilde{\rho}_{\ell
}L^{-1})}\right],  \label{eq:RRA6} 
\end{align}
\end{subequations}
where \eqref{eq:RRA4}-\eqref{eq:RRA7} are the two cases considering real roots
and the remaining equations \eqref{eq:RRA2}-\eqref{eq:RRA6} the four cases involving
one real and one complex root. Thus, the simplest cases with seasonal behavior in an MAR 
process are obtained with an MAR($1,1$) using \eqref{eq:RRA4} and \eqref{eq:RRA7} involving
only the zero and Nyquist frequency. Their respective partial fraction representations are given by
\begin{align}
y_{t} &= \frac{1}{1+\rho _{k}\tilde{\rho}_{\ell }}\left( \frac{-\rho _{k}L}{%
1+\rho _{k}L}+\frac{1}{1-\tilde{\rho}_{\ell }L^{-1}}\right) \varepsilon _{t},
\label{eq:par_fra_nar1_ar1_pi} \\
y_{t} &= \frac{1}{1+\rho _{k}\tilde{\rho}_{\ell }}\left( \frac{\rho _{k}L}{%
1-\rho _{k}L}+\frac{1}{1+\tilde{\rho}_{\ell }L^{-1}}\right) \varepsilon _{t}.
\label{eq:par_fra_ar1_nar1_pi}
\end{align}
In order to have MAR processes associated to a harmonic frequency we need 
to have lag or lead orders of at least two. As expressions rapidly become
larger, we illustrate such a situation for the MAR(1,2) process where the causal
polynomial has a root at the zero frequency and the noncausal polynomial
has a conjugate pair of roots. First, we define  
$\Delta^{NC}_{conj}(z^{-1}) := (1-\tilde{\rho}_{\ell }e^{-{\mathrm{i}}\tilde{\omega}%
_{\ell }}z^{-1})(1-\tilde{\rho}_{\ell }e^{{\mathrm{i}}\tilde{\omega}_{\ell
}}z^{-1}) = (1-2\cos \left( \tilde{\omega}%
_{\ell }\right)\tilde{\rho}_{\ell } z^{-1}+\tilde{\rho}_{\ell }^{2}z^{-2})$,
where the super- and subscript $NC$ and $conj$ indicate noncausal and conjugate respectively.
If we combine \eqref{eq:RRA2} and \eqref{eq:RRA3} with \eqref{eq:RA5}, we find:
\begin{equation} \label{eq:MAR(0,omega)}
\frac{1}{(1-\rho _{k}L)\Delta^{NC}_{conj}(L^{-1})} 
=
\frac{1}{(1-\rho _{k}L)}\left[ \frac{e^{-{\mathrm{i}}\tilde{\omega}_{\ell
}}/\left( e^{-{\mathrm{i}}\tilde{\omega}_{\ell }}-e^{{\mathrm{i}}\tilde{%
\omega}_{\ell }}\right) }{\left( 1-\tilde{\rho}_{\ell }e^{-{\mathrm{i}}%
\tilde{\omega}_{\ell }}L^{-1}\right) }+\frac{e^{{\mathrm{i}}\tilde{\omega}%
_{\ell }}/\left( e^{{\mathrm{i}}\tilde{\omega}_{\ell }}-e^{-{\mathrm{i}}%
\tilde{\omega}_{\ell }}\right) }{\left( 1-\tilde{\rho}_{\ell }e^{{\mathrm{i}%
}\tilde{\omega}_{\ell }}L^{-1}\right) }\right],    
\end{equation}
which can be further rewritten as:
\begin{align*}
&\frac{e^{-{\mathrm{i}}\tilde{\omega}_{\ell }}}{\left( e^{-{\mathrm{i}}%
\tilde{\omega}_{\ell }}-e^{{\mathrm{i}}\tilde{\omega}_{\ell }}\right) }\left[
\frac{1}{(1-\rho _{k}\tilde{\rho}_{\ell }e^{-{\mathrm{i}}\tilde{\omega}%
_{\ell }})}\left\{ \frac{\rho _{k}L}{(1-\rho _{k}L)}+\frac{1}{\left( 1-%
\tilde{\rho}_{\ell }e^{-{\mathrm{i}}\tilde{\omega}_{\ell }}L^{-1}\right) }%
\right\} \right]
+ \\
&\frac{e^{{\mathrm{i}}\tilde{\omega}_{\ell }}}{\left( e^{{\mathrm{i}}%
\tilde{\omega}_{\ell }}-e^{-{\mathrm{i}}\tilde{\omega}_{\ell }}\right) }%
\left[ \frac{1}{(1-\rho _{k}\tilde{\rho}_{\ell }e^{{\mathrm{i}}\tilde{\omega}%
_{\ell }})}\left\{ \frac{\rho _{k}L}{(1-\rho _{k}L)}+\frac{1}{\left( 1-%
\tilde{\rho}_{\ell }e^{{\mathrm{i}}\tilde{\omega}_{\ell }}L^{-1}\right) }%
\right\} \right]. 
\end{align*}%
Hence, based on \eqref{eq:MAR(0,omega)} we obtain: 
\begin{align}
y_{t} &= \left( \frac{e^{-{\mathrm{i}}\tilde{\omega}_{\ell }}}{ (e^{-{\mathrm{i}%
}\tilde{\omega}_{\ell }}-e^{{\mathrm{i}}\tilde{\omega}_{\ell }})}%
\frac{1}{(1-\rho _{k}\tilde{\rho}_{\ell }e^{-{\mathrm{i}}\tilde{\omega}%
_{\ell }})}\frac{\rho _{k}L}{(1-\rho _{k}L)}  
+\frac{e^{-{\mathrm{i}}\tilde{\omega}_{\ell }}}{(e^{-{\mathrm{i}}%
\tilde{\omega}_{\ell }}-e^{{\mathrm{i}}\tilde{\omega}_{\ell }})}\frac{%
1}{(1-\rho _{k}\tilde{\rho}_{\ell }e^{-{\mathrm{i}}\tilde{\omega}_{\ell }})}%
\frac{1}{\left( 1-\tilde{\rho}_{\ell }e^{-{\mathrm{i}}\tilde{\omega}_{\ell
}}L^{-1}\right)} \right) \varepsilon_{t} \notag \\
&+ \left( \frac{e^{{\mathrm{i}}\tilde{\omega}_{\ell}}}{(e^{{\mathrm{i}}\tilde{\omega}_{\ell}}-e^{-{%
\mathrm{i}}\tilde{\omega}_{\ell}})}\frac{1}{(1-\rho_{k}\tilde{\rho}_{\ell}e^{{\mathrm{i}}%
\tilde{\omega}_{\ell}})}\frac{\rho_{k}L}{(1-\rho_{k}L)}  
+\frac{e^{{\mathrm{i}}\tilde{\omega}_{\ell}}}{(e^{{\mathrm{i}}\tilde{\omega}_{\ell}} - e^{-{%
\mathrm{i}}\tilde{\omega}_{\ell}})}\frac{1}{(1-\rho_{k}\tilde{\rho}_{\ell}e^{{\mathrm{i}}%
\tilde{\omega}_{\ell}})}\frac{1}{(1-\tilde{\rho}_{\ell}e^{{\mathrm{i}}\tilde{\omega}_{\ell
}}L^{-1})} \right) \varepsilon_{t}. \label{eq:MAR(0,omega)2}
\end{align}%
Similar results can be shown for other combinations of causal and noncausal roots,
which have been collected in Appendix B to conserve space. Combining these findings with
\eqref{eq:par_fra_nar1_ar1_pi}--\eqref{eq:par_fra_ar1_nar1_pi} and \eqref{eq:MAR(0,omega)2}, 
we can conclude that power in the spectrum of the
MAR is due to separate effects of the monomials associated to the inverse
roots of the factorization of $\phi(z)$ and $\varphi(z^{-1})$. For higher
order MAR models, we obtain equivalent results as the ones reported for
the conventional AR model by combining \eqref{eq:RA1} (which breaks down 
in the \eqref{eq:RA2}--\eqref{eq:RA5} cases) and \eqref{eq:RRA1}
(covering the \eqref{eq:RRA4}--\eqref{eq:RRA6} cases). The overall conclusion is that any MAR model
admits a partial fraction representation where the factor associated to
different frequencies can be isolated. Therefore, it is impossible that $%
\phi(z)\varphi(z^{-1})$ jointly induce a seasonal effect whenever $\phi
(z)$ and $\varphi(z^{-1})$ do not separately affect a seasonal frequency.

\begin{figure}[tbp]
\captionsetup[subfigure]{labelformat=empty}
\centering
\begin{adjustbox}{minipage=\linewidth,scale=0.93}
\begin{subfigure}[b]{\textwidth}
								\centering
                \includegraphics[width=\textwidth]{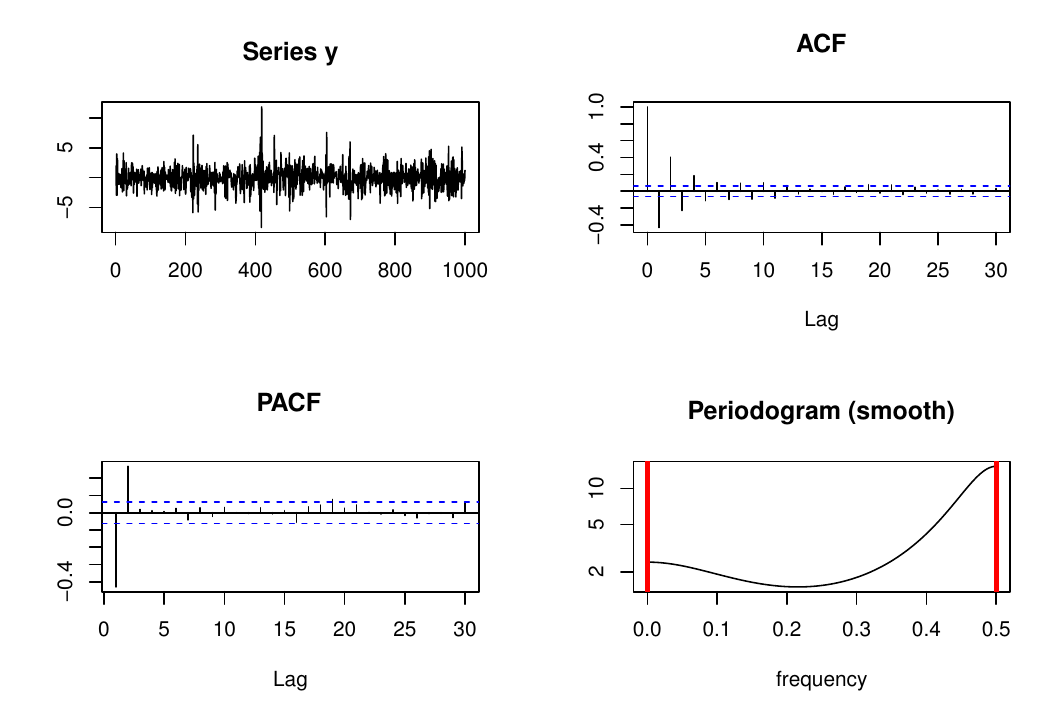}
                \caption{(a) Process with real-valued roots at the zero frequency and Nyquist frequency}
                \label{fig:ZeroNyquistMAR}
\end{subfigure}
\vskip\baselineskip
\begin{subfigure}[b]{\textwidth}
								\centering
                \includegraphics[width=\textwidth]{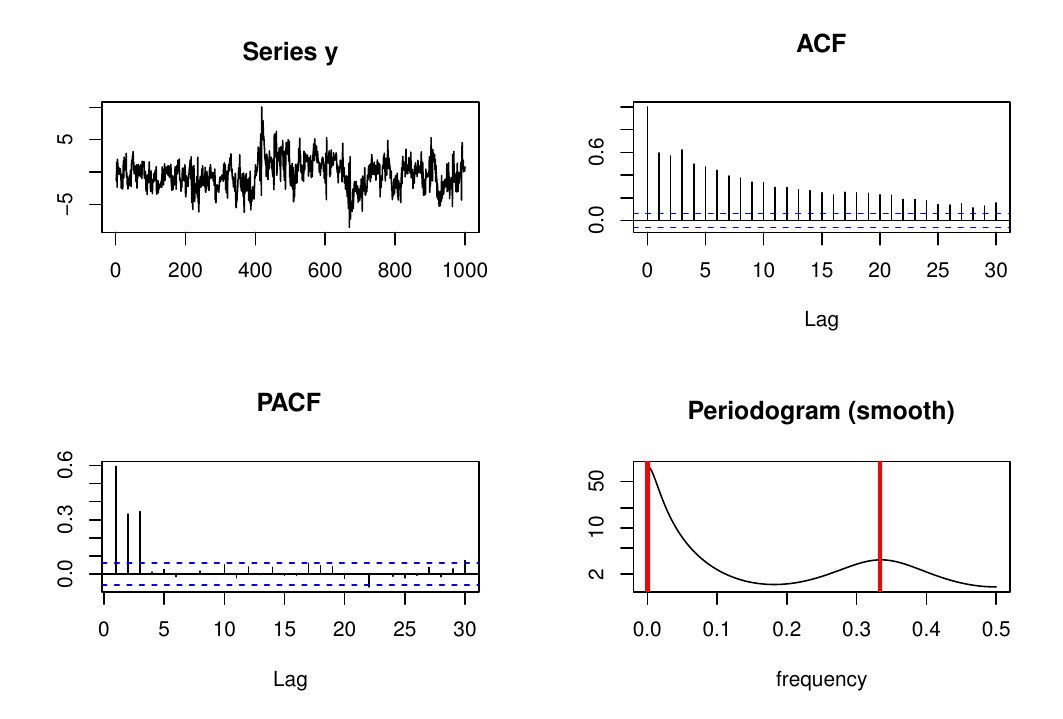}
                \caption{(b) Process with complex-valued roots at frequency $\frac{2}{3}\pi$
                and one real-valued root at the zero frequency}
                \label{fig:ZeroHarmonicMAR}
\end{subfigure} 
\end{adjustbox}
\caption{Simulated MAR processes including ACF, PACF and smoothed periodogram}
\label{fig:dataperiodMAR}
\begin{minipage}{\textwidth}\footnotesize
    Note: Frequencies have been mapped from $\left[ 0,\pi \right]$ to $\left[0,\frac{1}{2} \right]$, as the periodograms are displayed in the latter interval. The process in (a) represents a MAR(1,1) process with one root at the zero frequency (causal) and the other root at the Nyquist frequency $\pi$ (noncausal). Process (b) is an MAR(2,1) process that combines roots that appear in a pair of complex conjugates at frequency $\frac{2}{3}\pi$ (causal) and a root at the zero frequency (noncausal). The red lines in the periodograms highlight the expected peaks based on the chosen specifications.
\end{minipage}
\end{figure}

Figure \ref{fig:dataperiodMAR} collects time series plots and 
smoothed periodogram of simulated MAR time series, where the errors follow a $t(3,1)$ distribution. 
The root configurations of both processes are the same as in Figure \ref{fig:dataperiodAR}, with the difference that the roots have been divided over the causal and noncausal polynomials.
It is well-known that MAR processes can generate richer dynamics than their purely causal AR counterparts,
but we do not observe any clear differences in the periodograms of both figures. The spectrum peaks exactly
at the expected frequencies corresponding to the chosen roots and it does not make a difference whether
the roots belong to the backward- or forward-looking part of the model. This supports our theoretical
result that the seasonal effects are introduced through the causal and noncausal components separately: 
no new seasonal effects appear as a consequence of the multiplicative structure of the model. 

\subsubsection{Deterministic Seasonality}\label{sec:detseas}
Thus far, we have only investigated the role of stochastic seasonality in MAR models.
A deterministic seasonal component can be represented either as a linear combination of seasonal
dummy variables or as a linear combination of sine-cosine functions of various
frequencies \citep{Wei2006}. Using the latter method, we can extend \eqref {eq:mar} as follows: 
\begin{equation}  \label{eq:mar_seas}
\phi (L)\varphi (L^{-1})y_{t}= \sum _{h=0}^{\left\lfloor (S-1)/2 \right\rfloor}\left [\mu
_{h}^{\alpha }\cos \left (\frac{2\pi ht}{S}\right )+\mu _{h}^{\beta }\sin
\left (\frac{2\pi ht}{S}\right )\right ]+\mu _{S/2}\cos (\pi t)+\varepsilon
_{t},
\end{equation}
where we note that the model could be expanded even further by including other 
deterministics such as linear or polynomial time trends (in case these are deemed appropriate). The model in 
\eqref{eq:mar_seas} can be seen as an MAR model with exogenous regressors (MARX), which has been
studied in \cite{Hecqetal2020} and can analogously be estimated by approximate maximum likelihood.

The presence of deterministic seasonality can be detected using standard $t$- and $F$-tests on the $\mu$ coefficients. For the stochastic seasonality, we can find the seasonal frequency by collecting all the roots 
corresponding to the causal and noncausal polynomials and using the $\arctan2(a,b)$ function, 
which takes the real part $a$ and imaginary part $b$ of the root $z=a\pm b\mathrm{i}$ as argument
and returns the principal component. 

\begin{example}
Suppose we are given the polynomial $b(z) = 1 + 0.9z + 0.81z^2$,
which has roots $r_{1,2} = -0.556 \pm 0.962\mathrm{i}$ with corresponding
inverse roots $\alpha_{1,2} = -0.450 \pm 0.779\mathrm{i} $. We can obtain the
seasonal frequency as $\omega_{k} = \arctan2(-0.450,-0.779) = -2.618 \approx -\frac{5}{6}\pi$, 
which lies in the interval $(-\pi, \pi]$ as explained in Remark 1. 
\end{example}

\subsubsection{Comparison with Multiplicative Seasonal Model}
Practitioners often find that time series observations are not only related within periods but also between periods. For example, a monthly time series $\{y_{t}\}_{t \in \mathbb{Z}}$ can be temporally linked month-to-month, but also year-to-year. In the context of ARIMA models, this gives rise to the multiplicative seasonal ARIMA (SARIMA) model (see e.g., \citealp{Wei2006}), which makes these relations explicit. In a similar way,
a seasonal MAR (SMAR) model could be defined that explicitly allows for both the within- and 
between-period relationships to be potentially causal and noncausal. More specifically, we 
could define the SMAR($r,q$)$\times$($R,Q$)$_S$ model as
\begin{equation*}
\Phi(L^{S})\Psi(L^{-S})\phi(L)\varphi(L^{-1})y_{t} = \varepsilon_{t},
\end{equation*} 
where $\Phi(L^{S}) = 1 - \Phi_{1}L^{S} - ... - \Phi_{R}L^{RS}$ and $\Psi(L^{-S}) = 1 - \Psi_{1}L^{-S} - ... - \Psi_{Q}L^{-QS}$
are two seasonal polynomials with all roots strictly outside the unit circle and $S$ represents, as before, the integer-valued seasonal period. 

Note that our general framework implicitly covers this type of seasonal model. By setting $\Theta(z) = \phi(z)\Phi(z^{S})$ and $\Omega(z^{-1}) = \varphi(z^{-1})\Psi(z^{-S})$, which are polynomials of orders $r' := RS+r$ and $q' := QS+q$ respectively, the SMAR($r,q$)$\times$($R,Q$)$_S$ can be recast into a MAR($r',q'$) model with total autoregressive order $p'= r'+ q'$ and the results of Section \ref{sec:MAR} apply. Moreover, it could be argued that the term seasonal MAR is misleading, as we already explicitly allow for the presence of seasonal roots, as outlined in Section \ref{sec:seasonalroots}, in our definition of the MAR process. 

\section{Modelling Approach}\label{sec:EstModSel}
In this section, we study how the presence of seasonal roots affects the identification, estimation and model selection of MAR models. We show how the pseudo-causal model can be used to detect roots and explain how the presence of seasonal roots might simplify model selection.  

\subsection{Data Properties}\label{subsec:dataprop}
To remain as general as possible, we have only assumed that the error sequence $\{ \varepsilon_{t}\}_{t \in \mathbb{Z}}$ is $i.i.d.$ non-Gaussian. In the MAR parametric literature, we can distinguish two different strands: the finite-variance setting in which the rescaled $t$-distribution is a popular choice, and the heavy-tailed framework where the $\alpha$-stable distribution with $\alpha \in (0,2)$ is often employed. The chosen error distribution can be attributed to the type of empirical application: e.g. for inflation based on general price series, there is often no need to allow the error distribution to produce very extreme observations. This feature typically comes in play whenever one wants to model highly nonlinear patterns in the data, such as speculative bubbles or asymmetric cycles.

In the heavy-tailed framework, we often encounter that standard time-series measures such as the autocorrelation function lose their classical interpretation, but can still be employed (possibly in adapted form) as they are well-defined in the limit. Given our interest in associating roots in the MAR model to the correct frequency, we propose to estimate the power transfer function of the data $\{ y_{t} \}_{t=1}^{T}$ by means of the periodogram, defined as
\begin{equation}\label{eq:periodogram}
I_{n,y}(z) = r(n) \left| \sum_{t=1}^{T} y_{t} e^{-\mathrm{i}tz} \right|^{2}, \qquad z \in [-\pi, \pi],
\end{equation}
where $r(n) = n^{-1}$ in the finite variance, while $r(n) = n^{-2/\alpha}$ in the presence of an $\alpha$-stable distribution. This means that in the latter case, knowledge of the tail parameter $\alpha$ is required to compute an estimate of the power transfer function. Following \cite{Embrechtsetal1997}, we opt to use the self-normalized version of the periodogram given by
\begin{equation*}
\tilde{I}_{n,y} (z) = \frac{\left| \sum_{t=1}^{T} y_{t} e^{-\mathrm{i}tz} \right|^{2}}{\sum_{i=1}^{T} y_{t}^{2}}, \qquad z \in [-\pi, \pi],
\end{equation*}
where the dependence on $\alpha$ disappears as the term in the denominator grows at the same rate $r(n)$. In this way, we can infer whether there are any seasonal patterns present in the data before estimating MAR models.

The (self-normalized) periodogram can also help in detecting possible non-stationarity of the data. It is important to ensure that the series of interest is stationary, both at the zero and seasonal frequency. Whereas we do not elaborate on this point further in this paper, note that HEGY regression-based seasonal unit root tests \citep{Hyllebergetal1990} can be performed which also provide guidance on appropriate data transformations, if necessary.

\subsection{Approximate Maximum Likelihood Estimator}
To perform estimation of MAR models based on the principle of maximum likelihood, we follow the same procedure as \cite{LanneandSaikkonen2011}.
More specifically, we assume that $\varepsilon_{t}$ is non-Gaussian and that its distribution has a (Lebesgue) density 
$f_{\sigma}(x; \boldsymbol{\lambda}) = \sigma^{-1}f(\sigma^{-1}x; \boldsymbol{\lambda})$ satisfying the regularity conditions of \cite{Andrewsetal2006}, with the $d \times 1$ parameter vector $\boldsymbol{\lambda}$ collecting the distributional parameters in addition to the scale parameter $\sigma > 0$. We have an $r \times 1$ vector $\boldsymbol{\phi} = (\phi_{1},\ldots,\phi_{r})'$ and $q \times 1$ vector 
$\boldsymbol{\varphi} = (\varphi_{1},\ldots,\varphi_{q})'$ for the causal and noncausal coefficients, respectively. Their permissible parameter space of the autoregressive parameters is defined by the stationarity condition that the roots of both autoregressive polynomials lie strictly outside the unit circle. The approximate log-likelihood function for $\{ y_{t} \}_{t=1}^{T}$ is now given by
\begin{equation*}
l_{T}(\boldsymbol{\vartheta}) = \sum_{t=r+1}^{T-q} g_{t}(\boldsymbol{\vartheta}) = \sum_{t=r+1}^{T-q} \log f_{\sigma}(\phi(L)\varphi(L^{-1})y_{t};\boldsymbol{\lambda}),
\end{equation*}
where $\boldsymbol{\vartheta} = (\boldsymbol{\phi}', \boldsymbol{\varphi}', \sigma, \boldsymbol{\lambda}')'$ collects all autoregressive and distributional parameters. Maximizing $l_{T}(\boldsymbol{\vartheta})$ over permissible values of $\boldsymbol{\vartheta}$ gives an approximate maximum likelihood estimator (AMLE) of $\boldsymbol{\vartheta}$. Whereas the AMLE assumes a finite variance, simulation studies reveal that it also performs well in the infinite-variance case (see e.g., \citealp{Hecqetal2016}). However, to perform estimation we first need information on
the seasonal frequency $S$ and the autoregressive orders $(r, q)$ which are often unknown.

\subsection{Model Selection}
We adapt the model selection procedure of \cite{LanneandSaikkonen2011} to the context of seasonality by proposing the following steps:
\begin{itemize}
    \item[S1] In a first step, we propose to do an exploratory analysis in the spirit of Section \ref{subsec:dataprop}. That is, plot the data, identify possible seasonal behavior by means of the (self-normalized) periodogram and ensure stationarity of the data. 
    \item[S2] In a second step, the pseudo-causal model can 
    be estimated to identify the total autoregressive
    order and to confirm the presence of possible seasonal roots. By means of information criteria, correlograms, (partial) autocorrelation functions and Ljung-Box tests, it can be deduced at what frequencies a seasonal component is present and the lag order $p$ can be determined such that the residuals are free of serial correlation.
    \item[S3] The third step consists of estimating by AMLE all MAR($r,q$) with $p = r + q$ and selecting the model that maximizes the log-likelihood function at the estimated parameters.
\end{itemize}

Some further remarks are in place. In Step S1, it is important to take the features of MAR models into account. For unit root tests at the zero frequency, testing procedures are available in \cite{SaikkonenandSandberg2016} and \cite{Becetal2020}. Since a noncausal component can generate processes exhibiting conditional heteroskedasticity in direct time \citep{GourierouxandZakoian2017, FriesandZakoian2019}, we propose the extended HEGY tests in \cite{Cavaliereetal2019}. The test results provide guidance on how the original time series can be transformed in order to be stationary. In Step S2, we make use of the fact that any mixed causal-noncausal model can be expressed as a model with an autoregressive polynomial in lag operator $L$, which has all roots outside the unit circle. This model is second-order equivalent (SOE) and is often referred to as the pseudo-causal model.\footnote{In fact, multiple SOE models exist for a MAR model when not all roots are correctly assigned to the causal and noncausal part. Appendix C shows that the innovations corresponding to these models are all-pass (uncorrelated, but generally not independent).} The pseudo-causal model cannot only be used to determine the appropriate autoregressive orders, as \cite{FriesandZakoian2019} show 
that least squares estimation of the pseudo-causal representation ensures consistent identification of the roots of the MAR polynomial. Finally, the roots identified in the previous step can be used as starting values for the AMLE procedure in Step S3, where the final model is selected. Alternatively, if one does not want to use AMLE, it is possible to strictly rely on the OLS estimates and to perform an extreme residuals clustering approach to  find the strong form of the MAR \citep{FriesandZakoian2019}.
  
\subsection{Root Allocations}\label{subsec:rootall}
The results derived in Section \ref{sec:ARmodel} have important implications, because in theory the strong representation of the MAR process can be formed by obtaining the $p = r + q$ roots of the pseudo-causal model and assigning the correct $r$ roots to the causal polynomial and the remaining $q$ roots to the noncausal polynomial. 
%% Alternatively, this assignment can be used to provide appropriate starting values to the AMLE procedure, which also provides estimates of distributional parameters. 
In practice, however, the right allocation of roots to the causal and noncausal polynomial is unknown, as well as the total autoregressive order $p$ and the causal and noncausal orders $r$ and $q$. For this reason, we propose to estimate $p$ using pseudo-causal models. However, even for moderate autoregressive orders, it is quite cumbersome to try out all possible combinations of grouping $p$ roots in two groups of varying sizes $(r,q)$.  

\begin{example}\label{ex:roots}
Suppose we know that the true process is of order $p=4$, but the causal and noncausal orders $r$ and $q$ are unknown. Additionally, we have the four roots at our disposal. Now we can choose between the MAR(4,0), MAR(3,1), MAR(2,2), MAR(1,3) and MAR(0,4) model. Note that it is not straightforward how to assign the four roots to these models, except for the purely causal and noncausal case. For the MAR(3,1) and MAR(1,3), there are four possible root combinations, while this number is ${4 \choose 2} = 6$ for the MAR(2,2).
\end{example} 

Note that the root allocation problem outlined in Example \ref{ex:roots} simplifies when
a complex conjugate pair of roots is present in the pseudo-causal model. This pair has to be assigned \textit{jointly} to either the causal or noncausal part to ensure that the polynomials are still real-valued. To make the direct comparison, let us consider again the case $p=4$. It is straightforward to see that all models with both a causal and noncausal component now only have two possible root combinations. As processes with a total autoregressive order of $p > 4$ are relatively scarce, we argue that the presence of a complex conjugate pair of roots can simplify the estimation and model selection process for most relevant cases. On the difficult practical issue of picking starting values, \cite{Hecqvelasquez} have further discussed the choice of the roots for MAR models in a frequency domain framework; while \cite{Cubaddaetal2024} have proposed to rely on the simulated annealing algorithm to avoid getting trapped in local maxima.

\section{Monte Carlo Simulation}\label{sec:SimStudy}
Let us consider a MAR(1,2) process of the form
\begin{equation}\label{eq:DGP}
(1-\phi_{1}L)(1-2\cos(\omega_{k})\varphi_{1}L^{-1} + \varphi_{1}^{2}L^{-2})y_{t} = \varepsilon_{t},
\end{equation}
where the error term $\varepsilon_{t}$ follows a Student's $t(\nu,\sigma)$-distribution. 
We use this DGP to investigate two different topics: $(i)$ consistent estimation of the roots using the pseudo-causal model and $(ii)$ model selection. We consider different values for the autoregressive coefficients ($\phi_{1}, \varphi_{1}$), the frequency $\omega_{k}$ and the distributional parameters ($\nu, \sigma$) in the simulation
studies. All results are based on $10,000$ iterations.

\subsection{Pseudo-Causal Model}
To investigate whether we can consistently estimate the possibly complex-valued roots of the MAR model in the pseudo-causal representation, we set $\phi_{1} = 0.5$, $\varphi_{1} = 0.7$, $\omega_{k} = \frac{5}{6}\pi$, $\nu = 3$ and $\sigma = 1$ in \eqref{eq:DGP}.
It is easily seen that the root of the causal polynomial equal $r_{1} = 1/\alpha_{1} = 2$. The noncausal component contains a complex conjugate pair of roots such that the coefficients equal $b_{1} = -1.212$ and $b_{2} = -0.49$ respectively. Thus, we have to find the roots of the polynomial $b(z) = 1 + 1.212z + 0.49z^{2}$, which yields $r_{2,3} \approx -1.237\pm 0.714\mathrm{i}$. For different sample sizes $T \in \{ 100,200,500, 1000 \}$, we simulate 10,000 samples from the DGP. In every iteration, we estimate a causal AR(3) model by OLS, recover the roots and order them. We infer important information about the original process using the ``2-argument arctangent" function. It takes the real and imaginary part of the inverse roots as its first and second argument respectively, and provides the angle in radians in the interval $(-\pi,\pi]$ that an inverse root $\alpha_{k}$ makes with the positive real axis. For the DGP at hand, if we compute $\arctan2(\alpha_{k}^{R},\alpha_{k}^{I})$ based on $\alpha_{2}$ (or $\alpha_{3}$), we obtain the principal argument $\omega_{k} = \pm\frac{5}{6}\pi$. The inverse of the modulus, i.e. $[(\alpha_{k}^{R})^{2} + (\alpha_{k}^{I})^{2}]^{-1/2}$, based on $\alpha_{2,3}$ reveals that $\varphi_{1} = 0.7$. For the root at zero frequency, this yields  $\phi_{1} = 0.5$ as expected. 

 Table \ref{tab:roots} shows some properties
of the estimated roots: the average value $\mu_{\mbox{r}}$, average modulus $\mu_{\mbox{m}}$ and average inverse
modulus $\mu_{\mbox{im}}$ over all simulations. For $\mu_{\mbox{m}}$, the standard deviation is reported in parentheses.   
As expected, the results suggest that the roots can be consistently estimated, which is visible in two ways: 
the average (inverse) modulus comes closer to the true value and the standard deviation of the average modulus declines
as $T$ grows larger.    

\begin{table}[!h]
\centering
\begin{tabular}{lcccccccc}
\hline \hline
               & \multicolumn{2}{c}{$T=100$}  & \multicolumn{2}{c}{$T=200$} &  \multicolumn{2}{c}{$T=500$}  & \multicolumn{2}{c}{$T=1000$} \\ \hline 
               & $\alpha_{1}$  & $\alpha_{2,3}$  & $\alpha_{1}$ & $\alpha_{2,3}$  & $\alpha_{1}$   & $\alpha_{2,3}$  & $\alpha_{1}$  & $\alpha_{2,3}$      \\ \hline
$\mu_{\mbox{r}}$  & $2.373$    & $-1.211$  & $2.134$   & $-1.247$ & 2.053     & $-1.249$ & $2.026$    & $-1.243$  \\ 
               &          & $\pm 0.728\mathrm{i}$  & & $\pm 0.706\mathrm{i}$ & & $\pm 0.713\mathrm{i}$ & & $\pm 0.714\mathrm{i}$ \\ \hline
%$\mu_{ang}$   & 0.003    & -2.527        & 2.523         & 0.003   & -2.608        & 2.614         & $-3 \cdot 10^{-4}$    & -2.621        & 2.621         & $2 \cdot 10^{-16}$ & -2.620        & 2.620         \\ \hline
%$\mu_{ang\text{/}\pi}$  & $8 \cdot 10^{-4}$ & -0.804        & 0.803         & 0.001   & -0.830        & 0.832         & $-1 \cdot 10^{-4}$ & -0.834        & 0.834         & $5 \cdot 10^{-17}$ & -0.834        & 0.834         \\ \hline
$\mu_{\mbox{m}}$    & $2.777$    & $1.755$          & $2.154$     & $1.470$        & $2.054$       & $1.440$         & $2.026$    & $1.435$    \\ 
               & (16.224) & (5.362)        & (0.669)   & (0.298)      & (0.227)     & (0.079)       & (0.149)  & (0.052)  \\ \hline
$\mu_{\mbox{im}}$  & 0.463    & 0.671          & 0.483     & 0.690        & 0.492       & 0.696         & 0.496    & 0.698    \\ \hline \hline        
\end{tabular}
\caption{Mean of the roots ($\mu_{\mbox{r}}$), modulus ($\mu_{\mbox{m}}$) and inverse modulus ($\mu_{\mbox{im}}$)}
\label{tab:roots}
\end{table}

\begin{figure}[tbp]
\centering
\begin{adjustbox}{minipage=\linewidth,scale=0.75}
\begin{subfigure}[b]{0.475\textwidth}
								\centering
                \includegraphics[width=\textwidth]{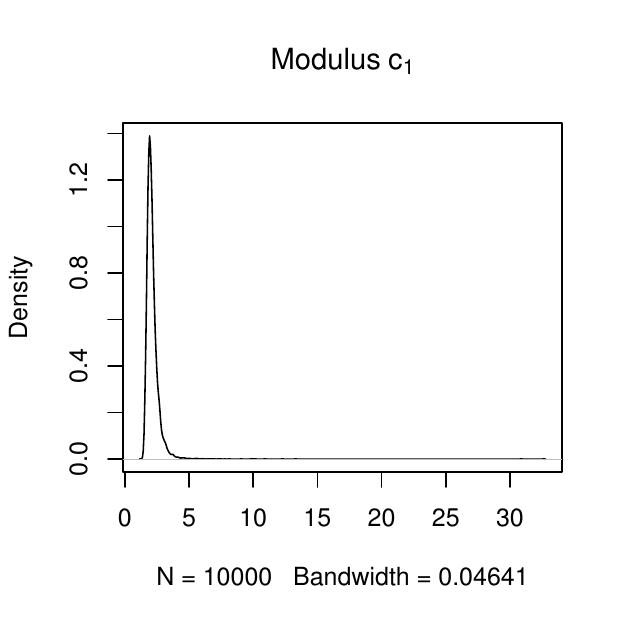}
                \caption{\footnotesize{Modulus root $\alpha_{1}$, $T=200$}}
                \label{fig:c1T200}
\end{subfigure} \hfill 
\begin{subfigure}[b]{0.475\textwidth}
                \includegraphics[scale=0.84]{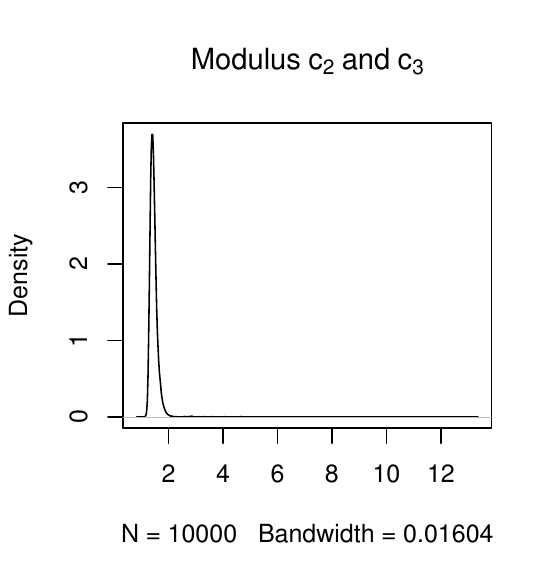}
                \caption{\footnotesize{Modulus roots $\alpha_{2},\alpha_{3}$, $T=200$}}
                \label{fig:c2c3T200}
        \end{subfigure}
\vskip\baselineskip
\begin{subfigure}[b]{0.475\textwidth}
								\centering
                \includegraphics[width=\textwidth]{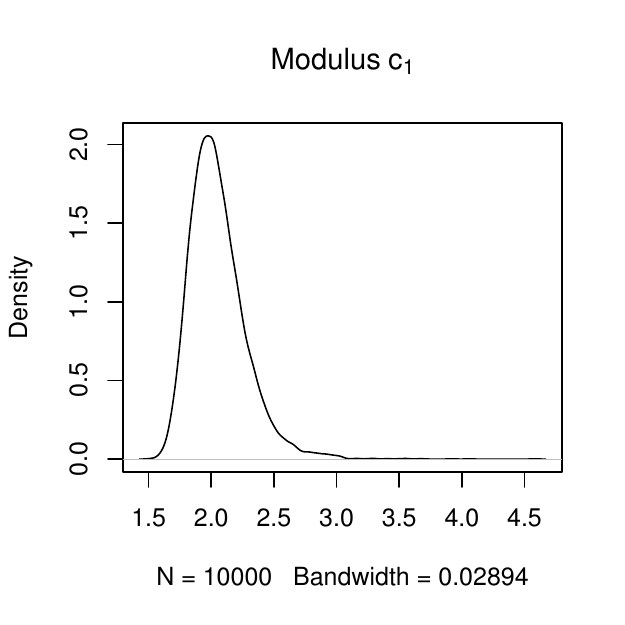}
                \caption{\footnotesize{Modulus root $\alpha_{1}$, $T=500$}}
                \label{fig:c1T500}
\end{subfigure} \hfill 
\begin{subfigure}[b]{0.475\textwidth}
                \includegraphics[width=\textwidth]{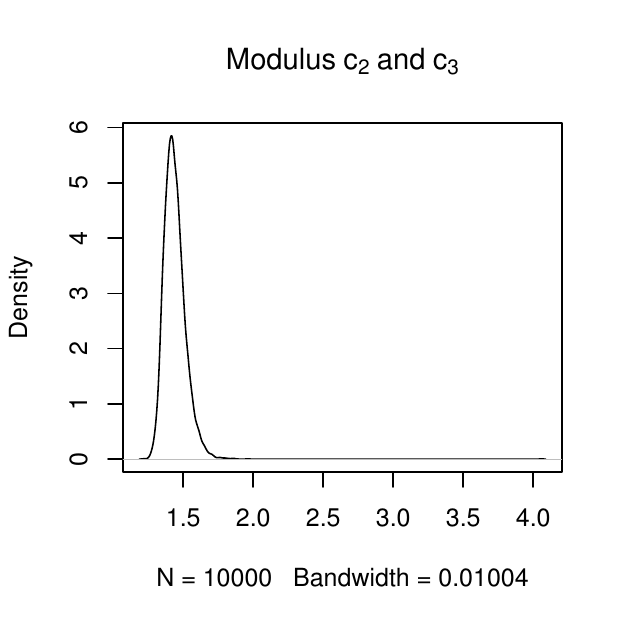}
                \caption{\footnotesize{Modulus roots $\alpha_{2},\alpha_{3}$, $T=500$}}
                \label{fig:c2c3T500}
        \end{subfigure}
\vskip\baselineskip
\begin{subfigure}[b]{0.475\textwidth}
                \includegraphics[width=\textwidth]{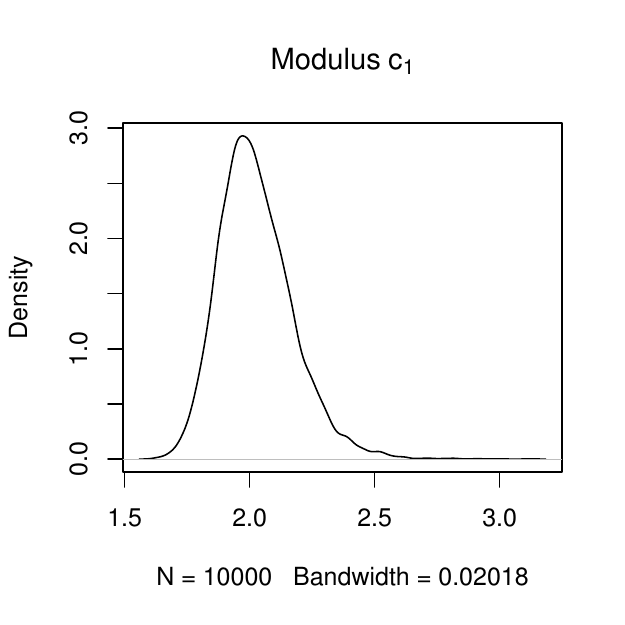}
                \caption{\footnotesize{Modulus root $\alpha_{1}$, $T=1000$}}
                \label{fig:c1T1000}
        \end{subfigure}
\quad 
\begin{subfigure}[b]{0.475\textwidth}
                \includegraphics[width=\textwidth]{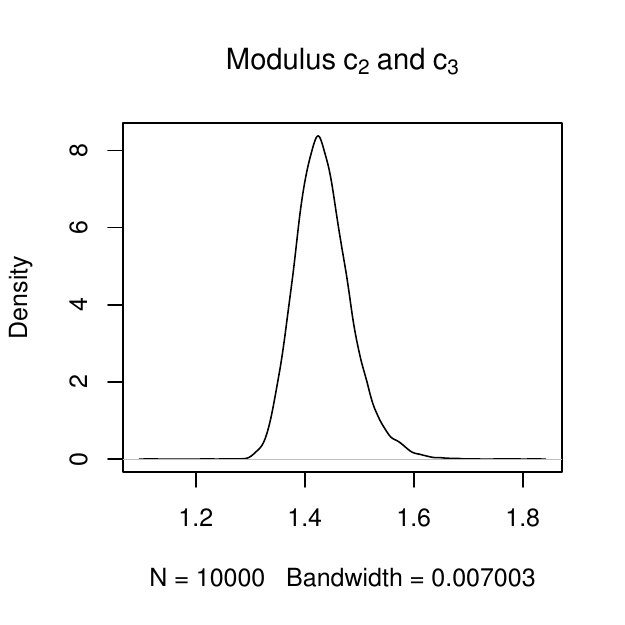}
                \caption{\footnotesize{Modulus root $\alpha_{2},\alpha_{3}$, $T=1000$}}
                \label{fig:c2c3T1000}
        \end{subfigure}
\end{adjustbox}
\caption{Distribution of the roots for different sample sizes $T=200,500,1000$}
\label{fig:distroots}
\end{figure}

Figure \ref{fig:distroots}, which displays the empirical distribution of the moduli of the roots, 
provides further support for this claim. For lower sample sizes, we see a larger right tail of the distribution, which 
reveals that the roots are not always accurately estimated. Note that this deviation from the true value can be 
in both directions: the fact that we observe a larger right tail is not surprising as the modulus is the absolute 
value of the roots. Two additional important observations have to be made. Firstly, we assume in this study that the 
total autoregressive order $p$ is known, whereas this is rarely the case in reality. This introduces another source of 
uncertainty, which might negatively affect the estimation of the roots. We decide to not further study this matter here, 
as it has been well-documented in the literature (see e.g. \citealp{LanneandSaikkonen2011} and \citealp{Hecqetal2020}). 
Secondly, we have to keep in mind that we can identify the roots in the pseudo-causal representation, but that we cannot know which roots belong to the causal and noncausal parts. As discussed in Section \ref{subsec:rootall}, this poses issues when we want to model the seasonality in the MAR directly. To prevent inaccurate estimation results, 
it is therefore important to ensure that pairs of complex conjugates are not split over the causal and noncausal polynomials in case they are used as initial values. Moreover, to circumvent problems of bi-modality (see e.g. \citealp{Hecqetal2016}), one could consider performing a grid search over starting values in the AML procedure \citep{Becetal2020}. To make this computationally feasible, the results in the pseudo-causal model provide guidance to what values should be considered in the grid. Alternatively, algorithms such as simulated annealing could be applied \citep{Cubaddaetal2024}. 
 
\subsection{Approximate Maximum Likelihood}
The previous simulations investigate whether the roots of the MAR process can be recovered by means of the pseudo-causal model by assuming that the total autoregressive order $p$ is known to the practitioner. If $p$ is unknown, this order can be determined quite adequately using diagnostic tests \citep{LanneandSaikkonen2011} and information criteria \citep{Hecqetal2016}. However, it is generally more challenging to find the corresponding causal and noncausal orders $r$ and $q$, as it requires the comparison of multiple non-nested models that have $p=r+q$. If treated correctly, the presence of complex conjugate root pairs simplifies model selection. This simulation study investigates the sensitivity of the AMLE selection procedure to seasonal roots appearing in pairs. 

More specifically, we consider the DGP in \eqref{eq:DGP} where we set $\phi_{1} = 0$, $\varphi_{1} \in \{0.3, 0.5, 0.7 \}$, $\omega_{k} = \frac{5}{6}\pi$, $\nu = 3$ and $\sigma = 1$. This means that the true process is a purely noncausal AR(2). We proceed as follows. On data simulated from the DGP, we estimate a  pseudo-causal model of order two. The obtained roots will be used as starting values for the candidate MAR models that are estimated by AMLE. In addition to the MAR(2,0) and MAR(0,2), we also consider the MAR(1,1) which represents the case in which we naively fail to supply the pair obtained from the pseudo-causal representation to a single polynomial. The model we select is the one that maximizes the log-likelihood at the estimated parameter values.

\begin{table}[H]
\centering
\begin{tabular}{l|ccc|ccc|ccc}
%\hline \hline
         & \multicolumn{3}{c|}{$\varphi_{1} = 0.3$} & \multicolumn{3}{c|}{$\varphi_{1} = 0.5$} & \multicolumn{3}{c}{$\varphi_{1} = 0.7$} \\ \cline{1-10}
         & $(2,0)$     & $(1,1)$     & $\textbf{(0,2)}$     & $(2,0)$     & $(1,1)$     & $\textbf{(0,2)}$     & $(2,0)$     & $(1,1)$     & $\textbf{(0,2)}$     \\ \hline
$T=100$  & $4.58\%$ & $33.06\%$ & $62.36\%$ &  $4.85\%$  & $8.64\%$             & $86.51\%$   & $5.30\%$  & $0.87\%$  & $93.83\%$            \\
$T=200$  & $0.53\%$  & $25.95\%$ & $73.52\%$ &  $0.57\%$  & $2.20\%$            & $97.23\%$ & $0.54\%$ & $0.06\%$   & $99.40\%$    \\
$T=500$  & $0.00\%$  & $11.03\%$ & $88.97\%$ & $0.00\%$ & $0.07\%$ & $99.93\%$ &  $0.00\%$           & $0.00\%$            & $100.00\%$            \\
$T=1000$ & $0.00\%$  & $2.69\%$  & $97.31\%$ &  $0.00\%$ & $0.00\%$ & $100.00\%$ & $0.00\%$  & $0.00\%$  & $100.00\%$      \\ \hline \hline     
\end{tabular}
\caption{Percentages with which the models are selected based on the highest log-likelihood.}
\label{tab:AMLE}
\end{table}

Table \ref{tab:AMLE} displays the selection of models for the different scenarios. Interestingly, we find that the mixed specification is a larger competitor to the true MAR(0,2) model than the purely causal alternative for $\varphi_{1} \in \{ 0.3, 0.5 \}$. Thus, the model selection procedure appears more proficient in detecting noncausality than recognizing that the process is based on a pair of complex conjugate roots, when the overall signal (as measured by $\varphi_{1}$) is relatively weaker. As expected, the selection becomes more accurate when sample size $T$ and the value of $\varphi_{1}$ grows. Since AML estimation typically requires starting values for the coefficients and not directly the roots, we decided to supply inverse of the modulus of the roots for both polynomials in the mixed specification. Although not directly obvious, it is of course possible to include the estimation results of the pseudo-causal model differently in the MAR(1,1) specification. For example, one could supply the inverse of the complex-valued root or supply only the reciprocal of the real part. We find that the results remain qualitatively similar in that situation. This emphasizes further that practitioners should interpret the pseudo-causal model's result carefully and rule out MAR alternatives that are not feasible a priori.  

\section{Empirical Illustrations}\label{sec:EmpApp}
In this section, we revisit existing empirical applications on COVID-19 and commodities data in the MAR literature. We interpret the presence of seasonal roots and explain how they affect the model selection procedure. 

\subsection{COVID-19 Data}
\begin{figure}[tbp]
    \centering
    \includegraphics[width=0.85\linewidth]{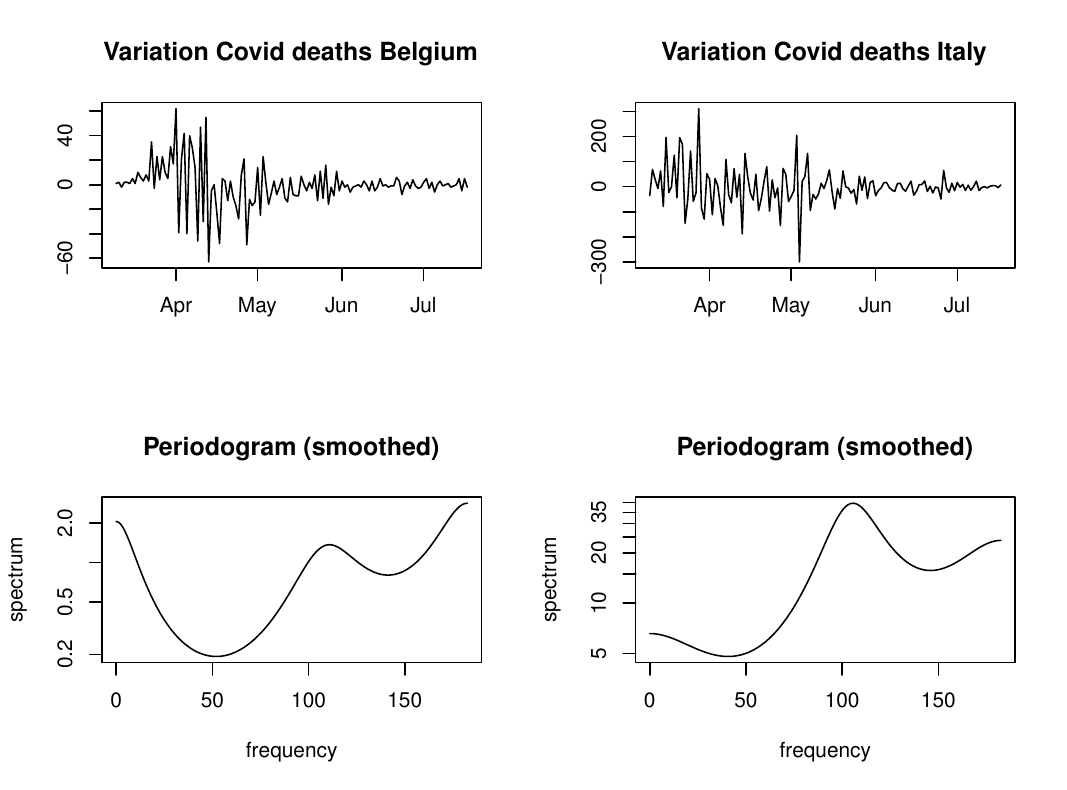}
    \caption{Variation in COVID-19 deaths for Belgium and Italy}
    \label{fig:empapp1}
\end{figure}

Similar to \cite{GiancateriniandHecq2025}, we consider the variation of daily COVID-19 deaths from March 10, 2020 to July 17, 2020, yielding $n = 130$ observations. In addition to Belgium, we also study the situation in Italy. The data is obtained from the World Health Organization (WHO) and both series are displayed together with their periodogram in Figure \ref{fig:empapp1}. The data exhibit large variations in the first days, which level out afterward. More importantly however, the zig-zag movement in Belgium during March-April 2020 resembles a possible seasonal bubble, as the amplitude of the series increases gradually over time. Thus, we expect the presence of at least one seasonal root (in particular, at the Nyquist frequency $\pi$). This increasing pattern is less pronounced for Italy and its periodogram looks different compared to Belgium in two ways. There does not appear to be a root at the zero and Nyquist frequency, but the peak in the middle (similar to Belgium) could point at the presence of complex roots. 

We start by estimating purely causal autoregressive models up to order $p_{max} = 14$ to identify the lag order which ensures that the residuals are free of serial correlation. Using the Bayesian Information Criterion (BIC), we find $p = 2$ for Italy and $p = 4$ for Belgium. Inspection of correlograms and additional diagnostic tests reveal the adequacy of these autoregressive orders. From the identified pseudo-causal AR models, we can deduce the possible presence of seasonality by computing the roots. 
The roots of the AR(2) for Italy are a pair of complex conjugates given by $-0.668 \pm 1.725\mathrm{i}$. For Belgium, we find two real-valued roots, i.e., $1.204$ and $-1.317$, which are associated to the zero and Nyquist frequency respectively, and a pair of complex conjugates, i.e., $-0.393\pm 1.315\mathrm{i}$. Applying the $\arctan2$-function to the complex roots of both Belgium at Italy reveal that the corresponding frequency equals $\frac{3}{5}\pi$. A Jarque-Bera test on the residuals of both models provides a $p$-value below $0.001$, which justifies distinguishing forward- and backward-looking behavior.

The roots obtained from the pseudo-causal models can be used to define starting values of the MAR($r,q$) models with $p = r + q$. Since we identify for both Belgium and Italy a pair of complex conjugates, we need to supply these jointly to either the causal or noncausal polynomial. This leads to an interesting scenario for Italy. Since \cite{FriesandZakoian2019} show that we can consistently estimate the roots of the MAR specification in the pseudo-causal model by least-squares, the MAR(1,1) can no longer be considered a viable option as it does not provide real-valued polynomials based on these roots. 
This reveals that the presence of seasonal roots can not only limit the possible combinations of starting values, but also the number of eligible models. The final model is selected as the one with the highest value of the log-likelihood at the estimated parameters, where we assume a Student's $t$-distribution for the error term with scale parameter $\sigma$ and degrees of freedom parameter $\nu$.\footnote{All MAR models include an intercept. For each estimated parameter, the corresponding standard error is provided in parentheses below.}  

The identification of MAR models results in a MAR($2,0$) for Italy and a MAR($2,2$) for Belgium. The purely causal model selected for Italy is given by
\begin{equation}
\left(
1 + \underset{(0.049)}{0.337}L + \underset{(0.049)}{0.335}L^{2}\right) 
(y_{t} + \underset{(3.270)}{5.099})  = e_{1,t},  
\label{Covid_model_Italy}
\end{equation}
with estimated scale $\widehat{\sigma} = 28.076$ and degrees of freedom $\widehat{\nu} = 1.677$. The low value of the degrees of freedom parameter highlights once again that a deviation of Gaussianity is appropriate, even though a causal model is selected. The evidence for the MAR(2,0) is quite convincing, given the difference of $15.277$ in log-likelihood value in favor of the causal specification ($-691.751$ versus $-707.028$). Interestingly, estimation of a MAR(1,1) using the starting values of the pseudo-causal model leads to a model with a log-likelihood value that lies in between the two pure specifications ($-705.241$). Thus, despite providing implausible starting values in the AMLE procedure, it still converges and delivers a model that is not strictly inferior to the other candidate models. This might be a small-sample issue, but practitioners are advised to carefully interpret the results of the pseudo-causal model when performing model selection in the next step.  

For Belgium we identify the following MAR(2,2) model
\begin{equation}
\left(1+\underset{(0.024)}{0.466}L + \underset{(0.024)}{0.585}L^{2} \right)
\left(
1- \underset{(0.023)}{0.080}L^{-1} - \underset{(0.023)}{0.604}L^{-2}\right) 
(y_{t} + \underset{(0.514)}{1.044})  = e_{2,t},  
\label{Covid_model_Belgium}
\end{equation}
with estimated scale $\widehat{\sigma} = 4.228$, degrees of freedom $\widehat{\nu} = 1.179$ and where the roots of the causal part represent a pair of complex conjugates. The inverse roots are of the form $\alpha _{R}\pm \mathrm{i}\alpha _{I}=$ $-0.398\pm \mathrm{i}1.245$, with
modulus $\left( \alpha _{R}^{2}+\alpha _{I}^{2}\right) ^{1/2}=\left( \left[-0.398\right] ^{2}+\left[ 1.245\right] ^{2}\right) ^{1/2} = 1.307$. The computation of $\arctan2(\alpha _{R},\alpha _{I})$ reveals that the polynomial is associated to frequency $1.880$, corresponding to oscillations that complete a full cycle every $2\pi /1.880$
periods (days). Therefore, it is possible to factorize this polynomial as
\begin{equation*}
\left( 1 + 0.466L + 0.585 L^{2} \right) = \left( 1-\left[ \frac{1}{1.307}%
\right] e^{-\mathrm{i}1.880}L\right) \left( 1-\left[ \frac{1}{1.307}\right]
e^{\mathrm{i}1.880}L\right). 
\end{equation*}
This part of the model explains cyclical or oscillating behavior of the time series after it reaches its highest value. The noncausal part correspond to the zero and Nyquist frequency with
the following factorization
\begin{equation*}
\left( 1-0.080L^{-1}-0.604L^{-2}\right) = \left( 1-\left[ \frac{1}{1.222}%
\right] L^{-1} \right) \left( 1+\left[ \frac{1}{1.355}\right] L^{-1} \right),
%\label{causal_part}
\end{equation*}
where the first factor associated to the zero frequency dominates over the second term related to the Nyquist frequency, due to it larger coefficient in absolute terms ($0.818$ compared to $0.738$, respectively). The first term explains the initial increasing behavior of the time series, while the latter term is responsible for the zig-zag behavior that follows and resembles a seasonal bubble.

\begin{figure}[tbp]
    \centering
    \includegraphics[width=0.85\linewidth]{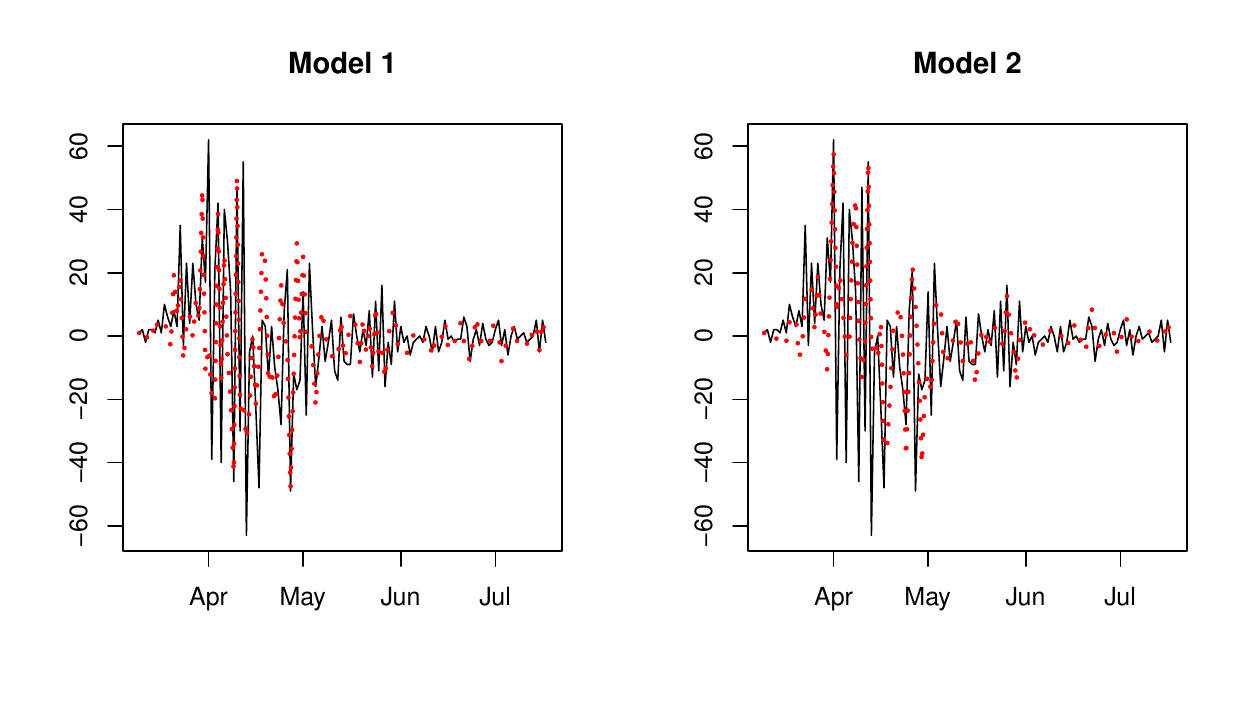}
    \caption{The fit of two competing MAR(2,2) models for Belgium (red dashed)}
    \label{fig:empapp2}
\end{figure}

An important remark is in place. The assignment of the roots to the causal and noncausal polynomial is crucial to identify the model with the highest log-likelihood value. The reported MAR(2,2) results are obtained by supplying the pair of complex conjugates to the causal polynomial and the two real-valued roots to the noncausal polynomial. If we allocate the roots the other way around, we instead obtain
\begin{equation*}
\left(1-\underset{(0.034)}{0.337}L - \underset{(0.034)}{0.538}L^{2} \right)
\left(
1 + \underset{(0.034)}{0.519}L^{-1} + \underset{(0.034)}{0.421}L^{-2}\right) 
(y_{t} + \underset{(0.636)}{0.562})  = e_{3,t},      
\end{equation*}
with $\widehat{\sigma} = 5.237$ and $\widehat{\nu} = 1.328$. Figure \ref{fig:empapp2} shows the fit (in dashed red) of the original and newly estimated model in the left and right panel, respectively. It can be seen that the original MAR(2,2) is able to capture the zig-zag behavior at the beginning of the series much better, as it has the root at the Nyquist frequency in the noncausal polynomial. The alternative specification is able to capture the upwards swings, but does a poor job in fitting the negative peaks. Another way to establish the superiority of the first model is by comparing log-likelihood values.
The new model has a log-likelihood value of $-491.895$, which is substantially lower than the value $-478.697$ for the previously identified model. This result emphasizes once again the danger of identifying local instead of global maxima, which can be circumvented by performing a grid search over starting values \citep{Becetal2020} or applying simulated annealing \citep{Cubaddaetal2024}.  

\subsection{Soybean Price}
\begin{figure}[tbp]
    \centering
    \includegraphics[width=\linewidth]{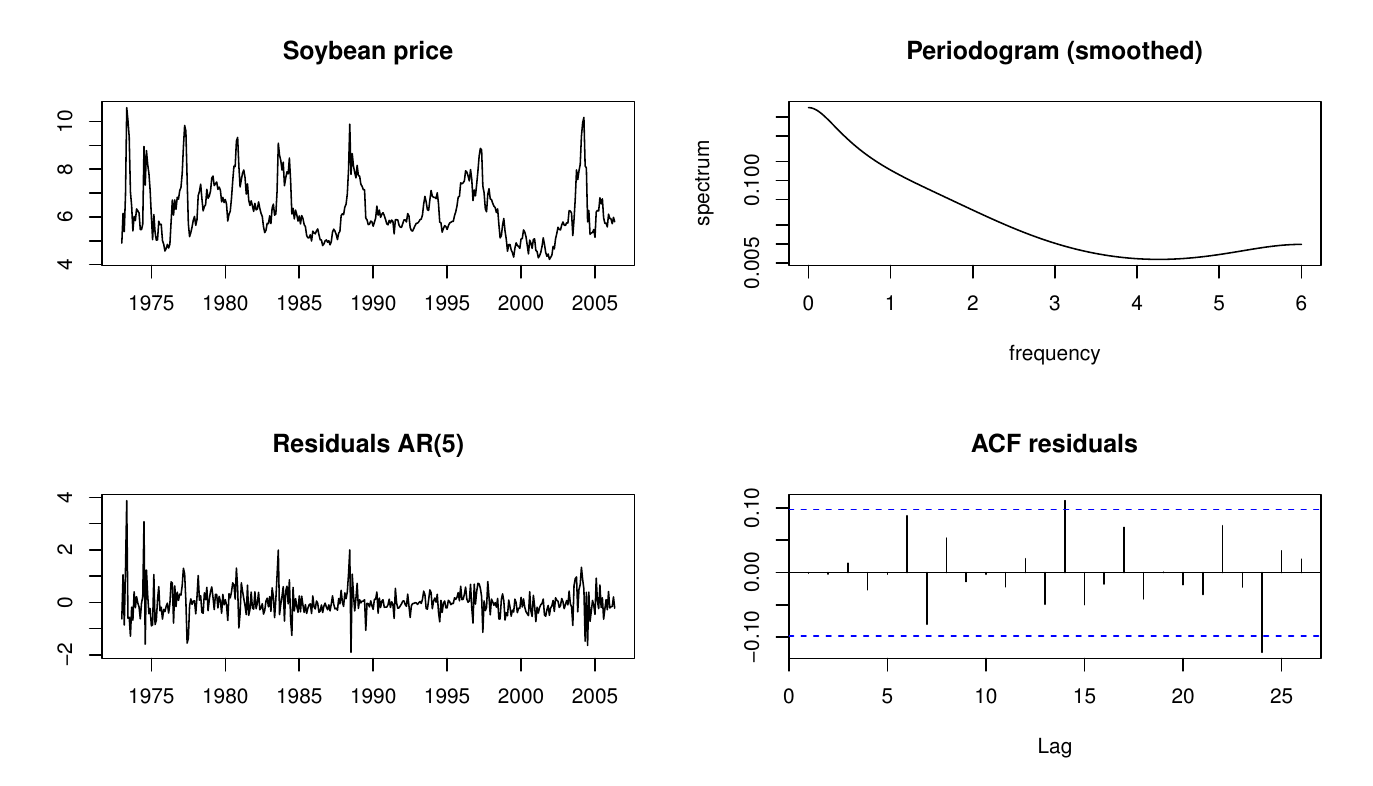}
    \caption{Soybean prices with periodogram (top row), residuals AR(5) with ACF (bottom row)}
    \label{fig:empapp3}
\end{figure}

We now focus on a financial series studied in \cite{FriesandZakoian2019}, the monthly soybean price measured in USD/bushel from January 1973 to May 2006.\footnote{\url{https://www.macrotrends.net/2531/soybean-prices-historical-chart-data}.} The time series, displayed in the top-left panel of Figure \ref{fig:empapp3}, shows recurrent episodes of local explosiveness, which makes it susceptible to both seasonality and noncausal autoregressive dynamics. The smoothed periodogram in the top-right panel reveals that we may expect roots at the zero frequency, at the Nyquist frequency $\pi$ and at least one pair of complex conjugates at frequency $\frac{k}{6}\pi$ for some $k \in \{ 1,2, \ldots, 5\}$. Similar to \cite{FriesandZakoian2019}, we find that an AR(5) is an appropriate pseudo-causal model based on BIC and additional diagnostic tests. The estimated model is given by
\begin{equation}\label{eq:soybeanpseudo}
\left( 1 - \underset{(0.050)}{0.954}L - \underset{(0.070)}{0.075}L^{2} + \underset{(0.069)}{0.280}L^{3} - \underset{(0.071)}{0.101}L^{4} - \underset{(0.054)}{0.017}L^{5} \right) ( y_{t} - \underset{(0.202)}{6.255}) = e_{1,t},    
\end{equation}
with an estimated error variance of $0.298$. The panels in the bottom row of Figure \ref{fig:empapp3} show the estimated model's residuals and its corresponding autocorrelation function. The residuals display peaks at most instances where the original series also peaked. This highlights the inability of a causal model to capture explosive, bubble-type behavior. The ACF confirms the absence of serial correlation.

The polynomial in \eqref{eq:soybeanpseudo} factorizes as $(1 - 0.852 L) (1+0.547 L) (1+0.125L) (1- 0.539e^{\mathrm{i}0.768}) (1- 0.539e^{-\mathrm{i}0.768})$,    
where the first term appeals to the zero frequency, the second and third term to the Nyquist frequency and the remaining two terms represent a pair of complex conjugates. We obtain the corresponding frequency by applying the $\arctan2$-function using the roots, i.e. $\arctan2(1.335,1.290) \approx 0.768$, which coincides with $\omega_{k} = \frac{1}{4}\pi$. However, since $S = 12$, the eligible frequencies are $\omega_{k} = 2\pi k/S = \pi k/6$, with $k \in \{ 1,2, \dots, 5 \}$. Note that for none of these values of $k$, we can obtain the frequency $\omega_{k} = \frac{1}{4}\pi$, as it is odd for a monthly process to complete a cycle every 8 months. The Jarque-Bera test on the residuals leads to a strong rejection of the null hypothesis of normality ($p$-value $< 0.001$) and thus we can look for signs of noncausality. Testing not only all MAR($r,q$) models with $r+q = 5$, but also applying all root combinations possible within a specific model, leads to the MAR(2,3) as the model with the highest log-likelihood. The AML estimation procedure yields the following result
\begin{equation}\label{eq:soybeanfinal}
\left(1+\underset{(0.024)}{0.305}L - \underset{(0.024)}{0.134}L^{2} \right) \left(1- \underset{(0.024)}{1.295}L^{-1} + \underset{(0.037)}{0.539}L^{-2} - \underset{(0.024)}{0.168}L^{-3} \right) (y_{t} - \underset{(0.018)}{0.517}) = e_{2,t},    
\end{equation}
with estimated scale $\widehat{\sigma} = 0.271$ and degrees of freedom parameter $\widehat{\nu} = 2.323$. 

Various remarks can be made. Firstly, factorizing the polynomials reveals that the causal part contains roots at the zero and Nyquist frequency, while the noncausal part has the pair of complex conjugate and a root at the zero frequency. Compared to the pseudo-causal representation, this means that one root at the Nyquist frequency has switched to a root at the zero frequency. This result seems more in line with the periodogram in Figure \ref{fig:empapp3}. Secondly, the obtained pair of complex conjugates in the MAR model equals $1.049 \pm 2.069\mathrm{i}$ and applying the $\arctan2$-function to these roots reveals that this coincides with $\omega_{k} = \frac{1}{3}\pi$, which holds for $k=2$. This fits better with our expectations for monthly data, as it means that a cycle is completed every 6 months. These findings differ compared to \citet[Table 5]{FriesandZakoian2019}, who rely on OLS estimation results of the pseudo-causal model and then use extreme residuals clustering to assign the roots to the causal and noncausal part of the model. In our model selection procedure,
we find one MAR(2,3) model based on a different starting value specification,
for which there is no switch of root type. However, this model obtains a log-likelihood value that is around eleven points lower than the selected model. Moreover, the roots differ substantially from those found in the pseudo-causal specification. Possible explanations are that the choice for the Student's $t$-distribution is inappropriate, or that the OLS estimates in the pseudo-causal model are not close enough to the true values for the data set at hand (see \citealp{FriesandZakoian2019}, Table 1, for simulation results on recovering the correct roots using OLS for different sample sizes and specifications of the error distributions). Based on the first two remarks made, we believe that our found results are sensible given the properties of the soybean price series.

An inspection of the fit of both the pseudo-causal AR(5) and MAR(2,3) model reveals that both are capturing the series quite well. As expected, we see that the AR(5) is underestimating the peaks and troughs in the time series. However, it has to be mentioned that the MAR(2,3) overestimates them at times, resulting in less large positive outliers but more negative ones. As the soybean prices display explosive episodes with possibly different rates of increase, considering an aggregation of noncausal models as proposed in \cite{GourierouxandZakoian2017} could be promising. However, estimation of such specifications requires further research, which is outside the scope of this paper.

\section{Conclusion}\label{sec:Conclusion}
This paper investigates seasonality in mixed causal-noncausal processes. Using the exponential form of inverse roots in combination with partial fraction decompositions, we show that the causal and noncausal parts are unable to generate new seasonal effects jointly in spite of the multiplicative structure of their polynomials. The seasonal effects can directly be isolated in the moving average representation of the process. Moreover, we find that seasonal roots can be identified using the pseudo-causal representation of the model and propose tools to study their exact behavior (e.g., modulus and frequency). In case of roots that appear in pairs of complex conjugates, we argue that the model selection for MAR model simplifies, as these roots have to be supplied jointly to the causal or noncausal polynomial. Monte Carlo simulations and two empirical illustrations support these findings and provide guidance to practitioners on how to interpret seasonality in MAR models.

%Future research could focus on identifying seasonality in the multivariate framework. In particular, it would be interesting to see whether a Jordan decomposition on the mixed causal-noncausal vector autoregressive model as formulated by \cite{GourierouxandJasiak2017} can isolate causal and noncausal seasonal roots. In that case, the semi-parametric GCov estimator could possibly consistently estimate these roots in one step. Another possible avenue is to consider (near-) unit root behavior at the zero and seasonal frequency components of the data, similar to \cite{DelBarrioCastroetal2019}. Not only the effects on estimation and model selection of MAR models could be of interest, but also the effects of aggregating MAR specifications as in \cite{Telg2024}.

\newpage

%\newpage
\bibliographystyle{Chicago-modified}
\bibliography{references}

\clearpage

\section*{Appendix}

\subsection*{A - Seasonal Autoregressive Processes}
In this paper, we focus on introducing the notion of seasonality in autoregressive processes. We contrast this with seasonal autoregressive processes, which are defined in the following way. Let $S$ be the seasonal periodicity, then we can write an autoregressive process of order $p \leq S$ as
\begin{equation*}
    a^{*}(L)y_{t} = \varepsilon^{*}_{t},
\end{equation*}
where $a^{*}(L) = 1 - \sum_{j=1}^{p} a^{*}_{j}L^{j}$. Following \cite{DelBarrioCastroetal2019}, we assume this polynomial can be factorized as
\begin{equation*}
    a(L) = \prod_{k=0}^{\left\lfloor S/2 \right\rfloor} \omega_{k}(L)^{h_{k}},
\end{equation*}
with $h_{k} \in \{ 0, 1 \}$ such that $p = \sum_{k=0}^{\left\lfloor S/2 \right\rfloor} f_{k}h_{k}$ with $f_{k}$ denoting the order of the real-valued polynomial $\omega_{k}(L)$.
Now there are three cases:
\begin{itemize}
    \item[(I)] $\omega_{0}(L) = (1- \alpha_{0}L)$ which associates inverse root $\alpha_{0}$ with frequency $\omega_{0} = 0$.
    \item[(II)] $\omega_{k}(L) = (1 - 2(\alpha_{k} \cos\omega_{k} - \beta_{k}\sin \omega_{k})L + (\alpha_{k}^{2} + \beta_{k}^{2})L^{2})$ which corresponds to conjugate seasonal frequencies $(\omega_{k}, 2\pi - \omega_{k})$, $\omega_{k} = \frac{2\pi k}{S}$ with associated parameters $\alpha_{k}$ and $\beta_{k}$, for $k = 1,\ldots, \left\lfloor \frac{(S-1)}{2} \right\rfloor$.
    \item[(III)] $\omega_{S/2}(L) = (1 + \alpha_{S/2}L)$ which associates inverse root $\alpha_{S/2}$ with the Nyquist frequency $\omega_{S/2} = \pi$ and is only defined in case $S$ is even.
\end{itemize}

\begin{example}
Suppose we have $S=12$ seasons per year, then we obtain $(i)$ the real-valued inverse roots $\alpha_{0}$ and $\alpha_{6}$ coming from $\omega_{0}(L)$ and $\omega_{6}(L)$, where both polynomials are of order one and correspond to the zero and Nyquist frequency $\pi$, respectively, and $(ii)$ the polynomials $\omega_{k}(L)$, $k=1,\ldots, 5$, which correspond to the frequencies $(\frac{1}{6}\pi, \frac{1}{3}\pi, \frac{1}{2}\pi, \frac{2}{3}\pi, \frac{5}{6}\pi)$, respectively. Since these appear in complex conjugate pairs, the order of the polynomials equals two. Thus, if all terms are contained within the autoregressive process, the order $p$ will indeed equal $S$. In the case at hand: $p = \sum_{k=0}^{6} f_{k}h_{k} = 2 \times 1 + 5 \times 2 = 12 $ (i.e., 2 polynomials of order 1 and 5 polynomials of order 2).
\end{example}

\subsection*{B - Partial Fraction Results for MAR(1,2) and MAR(2,2)}
We define  
$\Delta^{NC}_{conj} := (1-\tilde{\rho}_{\ell }e^{-{\mathrm{i}}\tilde{\omega}%
_{\ell }}L^{-1})(1-\tilde{\rho}_{\ell }e^{{\mathrm{i}}\tilde{\omega}_{\ell
}}L^{-1}) = (1-\tilde{\rho}_{\ell }2\cos \left( \tilde{\omega}%
_{\ell }\right) L^{-1}+\left( \tilde{\rho}_{\ell }\right) ^{2}L^{-2})$,
where the super- and subscript indicate noncausal and conjugate respectively. If we
consider the MAR(1,2) with a causal root at the Nyquist frequency and a complex
noncausal root, we get
\begin{equation}\label{eq:MAR(pi,omega)}
\frac{1}{(1+\rho _{k}L)\Delta^{NC}_{conj}} 
%=\frac{1}{(1+\rho _{k}L)(1-\tilde{\rho}_{\ell }e^{{\mathrm{i}}\tilde{\omega}%
%_{\ell }}L^{-1})(1-\tilde{\rho}_{\ell }e^{-{\mathrm{i}}\tilde{\omega}_{\ell
%}}L^{-1})}   \\
=\frac{1}{(1+\rho _{k}L)}\left[ \frac{e^{-{\mathrm{i}}\tilde{\omega}_{\ell
}}/\left( e^{-{\mathrm{i}}\tilde{\omega}_{\ell }}-e^{{\mathrm{i}}\tilde{%
\omega}_{\ell }}\right) }{\left( 1-\tilde{\rho}_{\ell }e^{-{\mathrm{i}}%
\tilde{\omega}_{\ell }}L^{-1}\right) }+\frac{e^{{\mathrm{i}}\tilde{\omega}%
_{\ell }}/\left( e^{{\mathrm{i}}\tilde{\omega}_{\ell }}-e^{-{\mathrm{i}}%
\tilde{\omega}_{\ell }}\right) }{\left( 1-\tilde{\rho}_{\ell }e^{-{\mathrm{i}%
}\tilde{\omega}_{\ell }}L^{-1}\right) }\right],  
\end{equation}
which can be rewritten as
\begin{align}
&\frac{e^{-{\mathrm{i}}\tilde{\omega}_{\ell }}}{\left( e^{-{\mathrm{i}}%
\tilde{\omega}_{\ell }}-e^{{\mathrm{i}}\tilde{\omega}_{\ell }}\right) }\left[
\frac{1}{(1+\rho _{k}\tilde{\rho}_{\ell }e^{-{\mathrm{i}}\tilde{\omega}%
_{\ell }})}\left\{ \frac{-\rho _{k}L}{(1+\rho _{k}L)}+\frac{1}{\left( 1-%
\tilde{\rho}_{\ell }e^{-{\mathrm{i}}\tilde{\omega}_{\ell }}L^{-1}\right) }%
\right\} \right] \notag \\
&+\frac{e^{{\mathrm{i}}\tilde{\omega}_{\ell }}}{\left( e^{{\mathrm{i}}%
\tilde{\omega}_{\ell }}-e^{-{\mathrm{i}}\tilde{\omega}_{\ell }}\right) }%
\left[ \frac{1}{(1+\rho _{k}\tilde{\rho}_{\ell }e^{{\mathrm{i}}\tilde{\omega}%
_{\ell }})}\left\{ \frac{-\rho _{k}L}{(1+\rho _{k}L)}+\frac{1}{\left( 1-%
\tilde{\rho}_{\ell }e^{{\mathrm{i}}\tilde{\omega}_{\ell }}L^{-1}\right) }%
\right\} \right].  \notag
\end{align}%
We also define $\Delta^{C}_{conj} := (1-\rho _{k}e^{-{\mathrm{i}}\omega _{k}}L)(1-\rho _{k}e^{{\mathrm{i}%
}\omega _{k}}L) = (1-\rho _{k}2\cos \left( \omega _{k}\right) L+\rho _{k}^{2}L^{2})$ for the case in which
the conjugate pair appears in the causal polynomial. For the MAR($2,1$) with such causal roots and a noncausal
root at the zero frequency, we combine \eqref{eq:RRA5} and \eqref{eq:RRA6} with \eqref{eq:RA5} to obtain:
\begin{equation}\label{eq:MAR(omega,0)}
\frac{1}{\Delta^{C}_{conj}(1-%
\tilde{\rho}_{\ell }L^{-1})} 
%&= \frac{1}{(1-\rho _{k}e^{-{\mathrm{i}}\omega
%_{k}}L)(1-\rho _{k}e^{{\mathrm{i}}\omega _{k}}L)(1-\tilde{\rho}_{\ell
%}L^{-1})}   \\
= \left[ \frac{e^{-{\mathrm{i}}\omega _{k}}/\left( e^{-{\mathrm{i}}\omega
_{k}}-e^{{\mathrm{i}}\omega _{k}}\right) }{\left( 1-\rho _{k}e^{-{\mathrm{i}}%
\omega _{k}}L\right) }+\frac{e^{{\mathrm{i}}\omega _{k}}/\left( e^{{\mathrm{i%
}}\omega _{k}}-e^{-{\mathrm{i}}\omega _{k}}\right) }{\left( 1-\rho _{k}e^{{%
\mathrm{i}}\omega _{k}}L\right) }\right] \frac{1}{(1-\rho _{k}L^{-1})} 
\end{equation}
which can be rewritten as
\begin{align*}
&\frac{e^{-{\mathrm{i}}\omega _{k}}}{\left( e^{-{\mathrm{i}}\omega
_{k}}-e^{{\mathrm{i}}\omega _{k}}\right) }\left[ \frac{1}{(1-\rho _{k}e^{-{%
\mathrm{i}}\omega _{k}}\tilde{\rho}_{\ell })}\left\{ \frac{\rho _{k}e^{-{%
\mathrm{i}}\omega _{k}}L}{\left( 1-\rho _{k}e^{-{\mathrm{i}}\omega
_{k}}L\right) }+\frac{1}{(1-\rho _{k}L^{-1})}\right\} \right]  \\
&+ \frac{e^{{\mathrm{i}}\tilde{\omega}_{\ell }}}{\left( e^{{\mathrm{i}}%
\tilde{\omega}_{\ell }}-e^{-{\mathrm{i}}\tilde{\omega}_{\ell }}\right) }%
\left[ \frac{1}{(1-\rho _{k}e^{{\mathrm{i}}\omega _{k}}\tilde{\rho}_{\ell })}%
\left\{ \frac{\rho _{k}e^{-{\mathrm{i}}\omega _{k}}L}{(\left( 1-\rho _{k}e^{-%
{\mathrm{i}}\omega _{k}}L\right) }+\frac{1}{(1-\rho _{k}L^{-1})}\right\} %
\right].  
\end{align*}
For the MAR(2,1) with a complex causal root and the noncausal root at the Nyquist frequency,
we obtain:
\begin{equation}\label{eq:MAR(omega,pi)}
\frac{1}{\Delta^{C}_{conj}(1+%
\tilde{\rho}_{\ell }L^{-1})}
%\frac{1}{(1-\rho _{k}e^{-{\mathrm{i}}\omega _{k}}L)(1-\rho _{k}e^{{\mathrm{i}%
%}\omega _{k}}L)(1+\tilde{\rho}_{\ell }L^{-1})} 
=\left[ \frac{e^{-{\mathrm{i%
}}\omega _{k}}/\left( e^{-{\mathrm{i}}\omega _{k}}-e^{{\mathrm{i}}\omega
_{k}}\right) }{\left( 1-\rho _{k}e^{-{\mathrm{i}}\omega _{k}}L\right) }+%
\frac{e^{{\mathrm{i}}\omega _{k}}/\left( e^{{\mathrm{i}}\omega _{k}}-e^{-{%
\mathrm{i}}\omega _{k}}\right) }{\left( 1-\rho _{k}e^{{\mathrm{i}}\omega
_{k}}L\right) }\right] \frac{1}{(1+\rho _{k}L^{-1})}
\end{equation}
which can be rewritten as
\begin{align*}
&\frac{e^{-{\mathrm{i}}\omega _{k}}}{\left( e^{-{\mathrm{i}}\omega
_{k}}-e^{{\mathrm{i}}\omega _{k}}\right) }\left[ \frac{1}{(1+\rho _{k}e^{-{%
\mathrm{i}}\omega _{k}}\tilde{\rho}_{\ell })}\left\{ \frac{\rho _{k}e^{-{%
\mathrm{i}}\omega _{k}}L}{\left( 1-\rho _{k}e^{-{\mathrm{i}}\omega
_{k}}L\right) }+\frac{1}{(1+\rho _{k}L^{-1})}\right\} \right] \\
&+ \frac{e^{{\mathrm{i}}\tilde{\omega}_{\ell }}}{\left( e^{{\mathrm{i}}%
\tilde{\omega}_{\ell }}-e^{-{\mathrm{i}}\tilde{\omega}_{\ell }}\right) }%
\left[ \frac{1}{(1+\rho _{k}e^{{\mathrm{i}}\omega _{k}}\tilde{\rho}_{\ell })}%
\left\{ \frac{\rho _{k}e^{-{\mathrm{i}}\omega _{k}}L}{(\left( 1-\rho _{k}e^{-%
{\mathrm{i}}\omega _{k}}L\right) }+\frac{1}{(1+\rho _{k}L^{-1})}\right\} %
\right].
\end{align*}%
These three models can thus be written in partial fraction decomposition. Firstly, for \eqref{eq:MAR(pi,omega)} we have: 
\begin{align}
y_{t} &= \frac{e^{-{\mathrm{i}}\tilde{\omega}_{\ell }}}{\left( e^{-{\mathrm{i%
}}\tilde{\omega}_{\ell }}-e^{{\mathrm{i}}\tilde{\omega}_{\ell }}\right) }%
\frac{1}{(1+\rho _{k}\tilde{\rho}_{\ell }e^{-{\mathrm{i}}\tilde{\omega}%
_{\ell }})}\frac{-\rho _{k}L}{(1+\rho _{k}L)}\varepsilon _{t}  \notag \\
&+ \frac{e^{-{\mathrm{i}}\tilde{\omega}_{\ell }}}{\left( e^{-{\mathrm{i}}%
\tilde{\omega}_{\ell }}-e^{{\mathrm{i}}\tilde{\omega}_{\ell }}\right) }\frac{%
1}{(1+\rho _{k}\tilde{\rho}_{\ell }e^{-{\mathrm{i}}\tilde{\omega}_{\ell }})}%
\frac{1}{\left( 1-\tilde{\rho}_{\ell }e^{-{\mathrm{i}}\tilde{\omega}_{\ell
}}L^{-1}\right) }\varepsilon _{t}  \label{eq:MAR(pi,omega)2} \\
&+ \frac{e^{{\mathrm{i}}\tilde{\omega}_{\ell }}}{\left( e^{{\mathrm{i}}%
\tilde{\omega}_{\ell }}-e^{-{\mathrm{i}}\tilde{\omega}_{\ell }}\right) }%
\frac{1}{(1+\rho _{k}\tilde{\rho}_{\ell }e^{{\mathrm{i}}\tilde{\omega}_{\ell
}})}\frac{-\rho _{k}L}{(1+\rho _{k}L)}\varepsilon _{t}  \notag \\
&+ \frac{e^{{\mathrm{i}}\tilde{\omega}_{\ell }}}{\left( e^{{\mathrm{i}}%
\tilde{\omega}_{\ell }}-e^{-{\mathrm{i}}\tilde{\omega}_{\ell }}\right) }%
\frac{1}{(1+\rho _{k}\tilde{\rho}_{\ell }e^{{\mathrm{i}}\tilde{\omega}_{\ell
}})}\frac{1}{\left( 1-\tilde{\rho}_{\ell }e^{{\mathrm{i}}\tilde{\omega}%
_{\ell }}L^{-1}\right) }\varepsilon _{t}.  \notag
\end{align}%
Secondly, \eqref{eq:MAR(omega,0)} becomes
\begin{align}
y_{t} &= \frac{e^{-{\mathrm{i}}\omega _{k}}}{\left( e^{-{\mathrm{i}}\omega
_{k}}-e^{{\mathrm{i}}\omega _{k}}\right) }\frac{1}{(1-\rho _{k}e^{-{\mathrm{i%
}}\omega _{k}}\tilde{\rho}_{\ell })}\frac{\rho _{k}e^{-{\mathrm{i}}\omega
_{k}}L}{\left( 1-\rho _{k}e^{-{\mathrm{i}}\omega _{k}}L\right) }\varepsilon
_{t}  \notag \\
&+ \frac{e^{-{\mathrm{i}}\omega _{k}}}{\left( e^{-{\mathrm{i}}\omega
_{k}}-e^{{\mathrm{i}}\omega _{k}}\right) }\frac{1}{(1-\rho _{k}e^{-{\mathrm{i%
}}\omega _{k}}\tilde{\rho}_{\ell })}\frac{1}{(1-\rho _{k}L^{-1})}\varepsilon
_{t}  \label{eq:MAR(omega,0)2} \\
&+ \frac{e^{{\mathrm{i}}\tilde{\omega}_{\ell }}}{\left( e^{{\mathrm{i}}%
\tilde{\omega}_{\ell }}-e^{-{\mathrm{i}}\tilde{\omega}_{\ell }}\right) }%
\frac{1}{(1-\rho _{k}e^{{\mathrm{i}}\omega _{k}}\tilde{\rho}_{\ell })}\frac{%
\rho _{k}e^{-{\mathrm{i}}\omega _{k}}L}{\left( 1-\rho _{k}e^{-{\mathrm{i}}%
\omega _{k}}L\right) }\varepsilon _{t}  \notag \\
&+ \frac{e^{{\mathrm{i}}\tilde{\omega}_{\ell }}}{\left( e^{{\mathrm{i}}%
\tilde{\omega}_{\ell }}-e^{-{\mathrm{i}}\tilde{\omega}_{\ell }}\right) }%
\frac{1}{(1-\rho _{k}e^{{\mathrm{i}}\omega _{k}}\tilde{\rho}_{\ell })}\frac{1%
}{(1-\rho _{k}L^{-1})}\varepsilon _{t}  \notag
\end{align}
And lastly, \eqref{eq:MAR(omega,pi)} yields:
\begin{align}
y_{t} &= \frac{e^{-{\mathrm{i}}\omega _{k}}}{\left( e^{-{\mathrm{i}}\omega
_{k}}-e^{{\mathrm{i}}\omega _{k}}\right) }\frac{1}{(1+\rho _{k}e^{-{\mathrm{i%
}}\omega _{k}}\tilde{\rho}_{\ell })}\frac{\rho _{k}e^{-{\mathrm{i}}\omega
_{k}}L}{\left( 1-\rho _{k}e^{-{\mathrm{i}}\omega _{k}}L\right) }\varepsilon
_{t}  \notag \\
&+ \frac{e^{-{\mathrm{i}}\omega _{k}}}{\left( e^{-{\mathrm{i}}\omega
_{k}}-e^{{\mathrm{i}}\omega _{k}}\right) }\frac{1}{(1+\rho _{k}e^{-{\mathrm{i%
}}\omega _{k}}\tilde{\rho}_{\ell })}\frac{1}{(1+\rho _{k}L^{-1})}\varepsilon
_{t}  \label{eq:MAR(omega,pi)2} \\
&+ \frac{e^{{\mathrm{i}}\tilde{\omega}_{\ell }}}{\left( e^{{\mathrm{i}}%
\tilde{\omega}_{\ell }}-e^{-{\mathrm{i}}\tilde{\omega}_{\ell }}\right) }%
\frac{1}{(1+\rho _{k}e^{{\mathrm{i}}\omega _{k}}\tilde{\rho}_{\ell })}\frac{%
\rho _{k}e^{-{\mathrm{i}}\omega _{k}}L}{\left( 1-\rho _{k}e^{-{\mathrm{i}}%
\omega _{k}}L\right) }\varepsilon _{t}  \notag \\
&+ \frac{e^{{\mathrm{i}}\tilde{\omega}_{\ell }}}{\left( e^{{\mathrm{i}}%
\tilde{\omega}_{\ell }}-e^{-{\mathrm{i}}\tilde{\omega}_{\ell }}\right) }%
\frac{1}{(1+\rho _{k}e^{{\mathrm{i}}\omega _{k}}\tilde{\rho}_{\ell })}\frac{1%
}{(1+\rho _{k}L^{-1})}\varepsilon _{t}.  \notag
\end{align}

For completeness, we study the only remaining case: both the causal and
noncausal polynomial have complex roots. We present results for an MAR($2,2$) in 
which we combine \eqref{eq:RA5} and \eqref{eq:RRA1} to get the desired result. 
In particular, we find that $\frac{1}{\Delta^{C}_{conj}\Delta^{NC}_{conj}}$ can be 
represented as
\begin{equation}\label{eq:MAR(2,2)}
%\end{equation}%
%\begin{equation*}
%\frac{1}{\left( 1-\rho _{k}e^{-{\mathrm{i}}\omega _{k}}L\right) \left(
%1-\rho _{k}e^{{\mathrm{i}}\omega _{k}}L\right) \left( 1-\tilde{\rho}_{\ell
%}e^{-{\mathrm{i}}\tilde{\omega}_{\ell }}L^{-1}\right) \left( 1-\tilde{\rho}%
%_{\ell }e^{{\mathrm{i}}\tilde{\omega}_{\ell }}L^{-1}\right) }=
%\end{equation*}%
%\begin{align*}
\left( \frac{e^{-{\mathrm{i}}\omega _{k}}/\left( e^{-{\mathrm{i}}\omega
_{k}}-e^{{\mathrm{i}}\omega _{k}}\right) }{\left( 1-\rho _{k}e^{-{\mathrm{i}}%
\omega _{k}}L\right) }+\frac{e^{{\mathrm{i}}\omega _{k}}/\left( e^{{\mathrm{i%
}}\omega _{k}}-e^{-{\mathrm{i}}\omega _{k}}\right) }{\left( 1-\rho _{k}e^{{%
\mathrm{i}}\omega _{k}}L\right) }\right) \left( \frac{e^{-{\mathrm{i}}\tilde{%
\omega}_{\ell }}/\left( e^{-{\mathrm{i}}\tilde{\omega}_{\ell }}-e^{{\mathrm{i%
}}\tilde{\omega}_{\ell }}\right) }{\left( 1-\tilde{\rho}_{\ell }e^{-{\mathrm{%
i}}\tilde{\omega}_{\ell }}L^{-1}\right) }+\frac{e^{{\mathrm{i}}\tilde{\omega}%
_{\ell }}/\left( e^{{\mathrm{i}}\tilde{\omega}_{\ell }}-e^{-{\mathrm{i}}%
\tilde{\omega}_{\ell }}\right) }{\left( 1-\tilde{\rho}_{\ell }e^{-{\mathrm{i}%
}\tilde{\omega}_{\ell }}L^{-1}\right) }\right),
\end{equation}%
which can be rewritten as
\begin{align*}
& \frac{e^{-{\mathrm{i}}\omega _{k}}}{\left( e^{-{\mathrm{i}}\omega _{k}}-e^{%
{\mathrm{i}}\omega _{k}}\right) }\frac{e^{-{\mathrm{i}}\tilde{\omega}_{\ell
}}}{\left( e^{-{\mathrm{i}}\tilde{\omega}_{\ell }}-e^{{\mathrm{i}}\tilde{%
\omega}_{\ell }}\right) }\left[ \frac{\rho _{k}e^{-{\mathrm{i}}\omega _{k}}L%
}{\left( 1-\rho _{k}e^{-{\mathrm{i}}\omega _{k}}L\right) }+\frac{1}{\left( 1-%
\tilde{\rho}_{\ell }e^{-{\mathrm{i}}\tilde{\omega}_{\ell }}L^{-1}\right) }%
\right] + \\
& \frac{e^{-{\mathrm{i}}\omega _{k}}}{\left( e^{-{\mathrm{i}}\omega _{k}}-e^{%
{\mathrm{i}}\omega _{k}}\right) }\frac{e^{{\mathrm{i}}\tilde{\omega}_{\ell }}%
}{\left( e^{{\mathrm{i}}\tilde{\omega}_{\ell }}-e^{-{\mathrm{i}}\tilde{\omega%
}_{\ell }}\right) }\left[ \frac{\rho _{k}e^{-{\mathrm{i}}\omega _{k}}L}{%
\left( 1-\rho _{k}e^{-{\mathrm{i}}\omega _{k}}L\right) }+\frac{1}{\left( 1-%
\tilde{\rho}_{\ell }e^{{\mathrm{i}}\tilde{\omega}_{\ell }}L^{-1}\right) }%
\right] + \\
& \frac{e^{{\mathrm{i}}\omega _{k}}}{\left( e^{{\mathrm{i}}\omega _{k}}-e^{-{%
\mathrm{i}}\omega _{k}}\right) }\frac{e^{-{\mathrm{i}}\tilde{\omega}_{\ell }}%
}{\left( e^{-{\mathrm{i}}\tilde{\omega}_{\ell }}-e^{{\mathrm{i}}\tilde{\omega%
}_{\ell }}\right) }\left[ \frac{\rho _{k}e^{{\mathrm{i}}\omega _{k}}L}{%
\left( 1-\rho _{k}e^{{\mathrm{i}}\omega _{k}}L\right) }+\frac{1}{\left( 1-%
\tilde{\rho}_{\ell }e^{-{\mathrm{i}}\tilde{\omega}_{\ell }}L^{-1}\right) }%
\right] + \\
& \frac{e^{{\mathrm{i}}\omega _{k}}}{\left( e^{{\mathrm{i}}\omega _{k}}-e^{-{%
\mathrm{i}}\omega _{k}}\right) }\frac{e^{{\mathrm{i}}\tilde{\omega}_{\ell }}%
}{\left( e^{{\mathrm{i}}\tilde{\omega}_{\ell }}-e^{-{\mathrm{i}}\tilde{\omega%
}_{\ell }}\right) }\left[ \frac{\rho _{k}e^{{\mathrm{i}}\omega _{k}}L}{%
\left( 1-\rho _{k}e^{{\mathrm{i}}\omega _{k}}L\right) }+\frac{1}{\left( 1-%
\tilde{\rho}_{\ell }e^{{\mathrm{i}}\tilde{\omega}_{\ell }}L^{-1}\right) }%
\right].
\end{align*}%
Thus, based on \eqref{eq:MAR(2,2)}, also the MAR(2,2) with complex causal and
noncausal roots admits a partial fraction representation given by
\begin{align}
& y_{t}=\frac{e^{-{\mathrm{i}}\omega _{k}}}{\left( e^{-{\mathrm{i}}\omega
_{k}}-e^{{\mathrm{i}}\omega _{k}}\right) }\frac{e^{-{\mathrm{i}}\tilde{\omega%
}_{\ell }}}{\left( e^{-{\mathrm{i}}\tilde{\omega}_{\ell }}-e^{{\mathrm{i}}%
\tilde{\omega}_{\ell }}\right) }\frac{\rho _{k}e^{-{\mathrm{i}}\omega _{k}}L%
}{\left( 1-\rho _{k}e^{-{\mathrm{i}}\omega _{k}}L\right) }\varepsilon _{t}
\label{eq:MAR(2,2)2} \\
&+ \frac{e^{-{\mathrm{i}}\omega _{k}}}{\left( e^{-{\mathrm{i}}\omega
_{k}}-e^{{\mathrm{i}}\omega _{k}}\right) }\frac{e^{-{\mathrm{i}}\tilde{\omega%
}_{\ell }}}{\left( e^{-{\mathrm{i}}\tilde{\omega}_{\ell }}-e^{{\mathrm{i}}%
\tilde{\omega}_{\ell }}\right) }\frac{1}{\left( 1-\tilde{\rho}_{\ell }e^{-{%
\mathrm{i}}\tilde{\omega}_{\ell }}L^{-1}\right) }\varepsilon _{t}  \notag \\
&+ \frac{e^{-{\mathrm{i}}\omega _{k}}}{\left( e^{-{\mathrm{i}}\omega
_{k}}-e^{{\mathrm{i}}\omega _{k}}\right) }\frac{e^{{\mathrm{i}}\tilde{\omega}%
_{\ell }}}{\left( e^{{\mathrm{i}}\tilde{\omega}_{\ell }}-e^{-{\mathrm{i}}%
\tilde{\omega}_{\ell }}\right) }\frac{\rho _{k}e^{-{\mathrm{i}}\omega _{k}}L%
}{\left( 1-\rho _{k}e^{-{\mathrm{i}}\omega _{k}}L\right) }\varepsilon _{t} 
\notag \\
&+ \frac{e^{-{\mathrm{i}}\omega _{k}}}{\left( e^{-{\mathrm{i}}\omega
_{k}}-e^{{\mathrm{i}}\omega _{k}}\right) }\frac{e^{{\mathrm{i}}\tilde{\omega}%
_{\ell }}}{\left( e^{{\mathrm{i}}\tilde{\omega}_{\ell }}-e^{-{\mathrm{i}}%
\tilde{\omega}_{\ell }}\right) }\frac{1}{\left( 1-\tilde{\rho}_{\ell }e^{{%
\mathrm{i}}\tilde{\omega}_{\ell }}L^{-1}\right) }\varepsilon _{t}  \notag \\
&+ \frac{e^{{\mathrm{i}}\omega _{k}}}{\left( e^{{\mathrm{i}}\omega _{k}}-e^{-%
{\mathrm{i}}\omega _{k}}\right) }\frac{e^{-{\mathrm{i}}\tilde{\omega}_{\ell
}}}{\left( e^{-{\mathrm{i}}\tilde{\omega}_{\ell }}-e^{{\mathrm{i}}\tilde{%
\omega}_{\ell }}\right) }\frac{\rho _{k}e^{{\mathrm{i}}\omega _{k}}L}{\left(
1-\rho _{k}e^{{\mathrm{i}}\omega _{k}}L\right) }\varepsilon _{t}  \notag \\
&+ \frac{e^{{\mathrm{i}}\omega _{k}}}{\left( e^{{\mathrm{i}}\omega _{k}}-e^{-%
{\mathrm{i}}\omega _{k}}\right) }\frac{e^{-{\mathrm{i}}\tilde{\omega}_{\ell
}}}{\left( e^{-{\mathrm{i}}\tilde{\omega}_{\ell }}-e^{{\mathrm{i}}\tilde{%
\omega}_{\ell }}\right) }\frac{1}{\left( 1-\tilde{\rho}_{\ell }e^{-{\mathrm{i%
}}\tilde{\omega}_{\ell }}L^{-1}\right) }\varepsilon _{t}  \notag \\
&+ \frac{e^{{\mathrm{i}}\omega _{k}}}{\left( e^{{\mathrm{i}}\omega _{k}}-e^{-%
{\mathrm{i}}\omega _{k}}\right) }\frac{e^{{\mathrm{i}}\tilde{\omega}_{\ell }}%
}{\left( e^{{\mathrm{i}}\tilde{\omega}_{\ell }}-e^{-{\mathrm{i}}\tilde{\omega%
}_{\ell }}\right) }\frac{\rho _{k}e^{{\mathrm{i}}\omega _{k}}L}{\left(
1-\rho _{k}e^{{\mathrm{i}}\omega _{k}}L\right) }\varepsilon _{t}  \notag \\
&+ \frac{e^{{\mathrm{i}}\omega _{k}}}{\left( e^{{\mathrm{i}}\omega _{k}}-e^{-%
{\mathrm{i}}\omega _{k}}\right) }\frac{e^{{\mathrm{i}}\tilde{\omega}_{\ell }}%
}{\left( e^{{\mathrm{i}}\tilde{\omega}_{\ell }}-e^{-{\mathrm{i}}\tilde{\omega%
}_{\ell }}\right) }\frac{1}{\left( 1-\tilde{\rho}_{\ell }e^{{\mathrm{i}}%
\tilde{\omega}_{\ell }}L^{-1}\right) }\varepsilon _{t}.  \notag
\end{align}%

\subsection*{C - All-Pass Representation}
Based on \eqref{eq:mar.alt}, \cite{Breidtetal2001} derive that
any MAR can be rewritten as a model with an autoregressive polynomial in lag operator
$L$, which has all roots outside the unit circle as in equation \eqref{eq:ar}.
This model is equivalent from a second-order perspective and is often referred to
as pseudo-causal model. Consider the two alternative formulations of the MAR given by
\begin{equation}
a(L)y_{t} = \epsilon_{t} \quad \text{ and } \quad \phi(L)\varphi(L^{-1})y_{t} = \varepsilon_{t},    
\end{equation}
which are \eqref{eq:mar.alt} and \eqref{eq:mar} respectively. Now assume the true process is \eqref{eq:mar.alt} and let $\varphi(z)$ be
the causal polynomial whose roots are reciprocals of those of $\varphi^{\ast}(z)$, which means that $\varphi(z)$ has the same coefficients as $\varphi(z^{-1})$. For the first formulation, we have
\begin{equation}\label{eq:soe}
a^{\ast}(L)y_{t} = \phi(L)\varphi(L)y_{t} = \varepsilon^{\ast}_{t},
\end{equation}
such that
\begin{equation*}
\varepsilon^{\ast}_{t} = \frac{\phi(L)\varphi(L)}{\phi(L)\varphi^{*}(L)}\epsilon_{t} = \frac{\varphi(L)}{-\varphi^{\ast}_{q} L^{q}\varphi(L^{-1})} \epsilon_{t} = \frac{\varphi(L)}{\varphi(L^{-1})} \tilde{\epsilon}_{t},
\end{equation*}
with $\tilde{\epsilon}_{t} = (-1/\varphi^{\ast}_{q})\epsilon_{t+q}$. Since $\{ \tilde{\epsilon}_{t} \}_{t \in \mathbb{Z}}$ is a rescaled 
and time-shifted version of $\{ \epsilon_{t} \}_{t \in \mathbb{Z}}$, it is still an $i.i.d$. sequence and thus
$\{\varepsilon^{\ast}_{t} \}_{t \in \mathbb{Z}}$ is a noncausal all-pass process of order $q$. If we choose for the formulation using lag and lead polynomial, i.e. \eqref{eq:mar}, we again obtain \eqref{eq:soe} but with error term
\begin{equation*}
\varepsilon^{\ast}_{t} = \frac{\phi(L)\varphi(L)}{\phi(L)\varphi(L^{-1})}\varepsilon_{t} = \frac{\varphi(L)}{\varphi(L^{-1})}\varepsilon_{t},    
\end{equation*}
which reveals that $\{\varepsilon^{\ast}_{t} \}_{t \in \mathbb{Z}}$ is once again a noncausal all-pass process of order $q$. For $\{\epsilon_{t} \}_{t \in \mathbb{Z}}$ and $\{ \varepsilon_{t} \}_{t \in \mathbb{Z}}$ $i.i.d.$ sequences with zero
mean and finite variance, $\{\varepsilon^{\ast}_{t} \}_{t \in \mathbb{Z}}$ is an uncorrelated sequence. When the error sequences are in addition Gaussian, we have that $\{ \varepsilon^{\ast}_{t} \}_{t \in \mathbb{Z}}$ is $i.i.d.$, which reveals why it is impossible to discriminate between
the pseudo-causal model and the MAR model, irrespective of its formulation. 
\cite{AndrewsandDavis2013} show that $\{\varepsilon^{\ast}_{t}\}_{t \in \mathbb{Z}}$ is an empirically uncorrelated all-pass sequence 
in case $\{\epsilon_{t}\}_{t \in \mathbb{Z}}$ or $\{ \varepsilon_{t} \}_{t \in \mathbb{Z}}$ are $i.i.d.$ sequences with infinite variance (for which the theoretical correlations do not exist).

Generally, wrongly distributed roots to either the causal or noncausal component
of a MAR process, based on either \eqref{eq:mar.alt} or \eqref{eq:mar}, result in all-pass representations. We characterize this finding only for the latter model formulation. Note that if our goal is to find the true process, i.e. the strong form, we might allocate the $p$ roots identified from the pseudo-causal model wrongly to the causal and noncausal component.
Let us consider the case where only parts of the roots are assigned correctly.
That is, suppose we have $\phi(z) = \phi_{1}(z)\phi_{2}(z)$ and $\varphi(z^{-1}) = \varphi_{1}(z^{-1})\varphi_{2}(z^{-1})$,
but only the roots in $\phi_{1}(z)$ and $\varphi_{1}(z^{-1})$ are allocated correctly. Then, we obtain
\begin{align*}
&\phi_{1}(L)\phi_{2}(L^{-1})\varphi_{1}(L^{-1})\varphi_{2}(L)y_{t} = \varepsilon^{*}_{t} \\
&\phi_{1}(L)\phi_{2}(L^{-1})\varphi_{1}(L^{-1})\varphi_{2}(L)[\phi(L)\varphi(L^{-1})]^{-1}\varepsilon_{t} = \varepsilon^{*}_{t},
\end{align*}
which leads to the expression
\begin{equation}\label{eq:MARallpass1}
\varepsilon^{*}_{t} = \frac{\phi_{1}(L)\phi_{2}(L^{-1})\varphi_{1}(L^{-1})\varphi_{2}(L)}{\phi_{1}(L)\phi_{2}(L)\varphi_{1}(L^{-1})\varphi_{2}(L^{-1})}\varepsilon_{t} = \frac{\phi_{2}(L^{-1})\varphi_{2}(L)}{\phi_{2}(L)\varphi_{2}(L^{-1})}\varepsilon_{t}.
\end{equation}
Let $\phi^{*}_{2}(z)$ be the polynomial whose roots are reciprocal to those of $\phi_{2}(z)$, and $\varphi^{*}_{2}(z^{-1})$ the polynomial
whose roots are reciprocal to those of $\varphi_{2}(z^{-1})$. Denote the orders of $\phi_{2}(z)$ and $\varphi_{2}(z^{-1})$ by $r_{2}$ and $q_{2}$ respectively, then we obtain
\begin{align*}
\phi^{*}_{2}(z) = -\phi_{2,r_{2}}^{-1}z^{r_{2}}\phi_{2}(z^{-1}), \\
\varphi^{*}_{2}(z^{-1}) = -\varphi_{2,q_{2}}^{-1}z^{-q_{2}}\varphi_{2}(z),
\end{align*}
with $\phi_{2,r_{2}}$ and $\varphi_{2,q_{2}}$ the coefficients corresponding to the term $z^{r_{2}}$ in $\phi_{2}(z)$ and $z^{-q_{2}}$ in $\varphi_{2}(z^{-1})$, respectively. Substituting these results in \eqref{eq:MARallpass1} yields
\begin{align*}
\varepsilon^{*}_{t} &= \frac{\left[-\phi_{2,r_{2}}L^{-r_{2}}\phi^{*}_{2}(L)\right]\left[-\varphi_{2,q_{2}}L^{q_{2}}\varphi^{*}_{2}(L^{-1})\right]}{\phi_{2}(L)\varphi_{2}(L^{-1})}\varepsilon_{t} \\
&= \frac{\phi^{*}_{2}(L)}{\phi_{2}(L)} \frac{\varphi^{*}_{2}(L^{-1})}{\varphi_{2}(L^{-1})}\left[\phi_{2,r_{2}}\varphi_{2,q_{2}}L^{q_{2} - r_{2}} \varepsilon_{t}\right] 
= \frac{\phi^{*}_{2}(L)}{\phi_{2}(L)} \frac{\varphi^{*}_{2}(L^{-1})}{\varphi_{2}(L^{-1})}\tilde{\varepsilon}_{t},
\end{align*}
Since $\{\tilde{\varepsilon}_{t}\}_{t \in \mathbb{Z}}$ is a rescaled and time-shifted version of $\{\varepsilon_{t}\}_{t \in \mathbb{Z}}$, it is an $i.i.d.$ sequence and thus we have that
$\{\varepsilon^{*}_{t}\}_{t \in \mathbb{Z}}$ is a mixed causal-noncausal all-pass sequence.

\subsection*{D - Figures}
\begin{figure}[H]
\centering
\begin{adjustbox}{minipage=\linewidth,scale=0.9}
\begin{subfigure}[b]{0.475\textwidth}
								\centering
                \includegraphics[width=\textwidth]{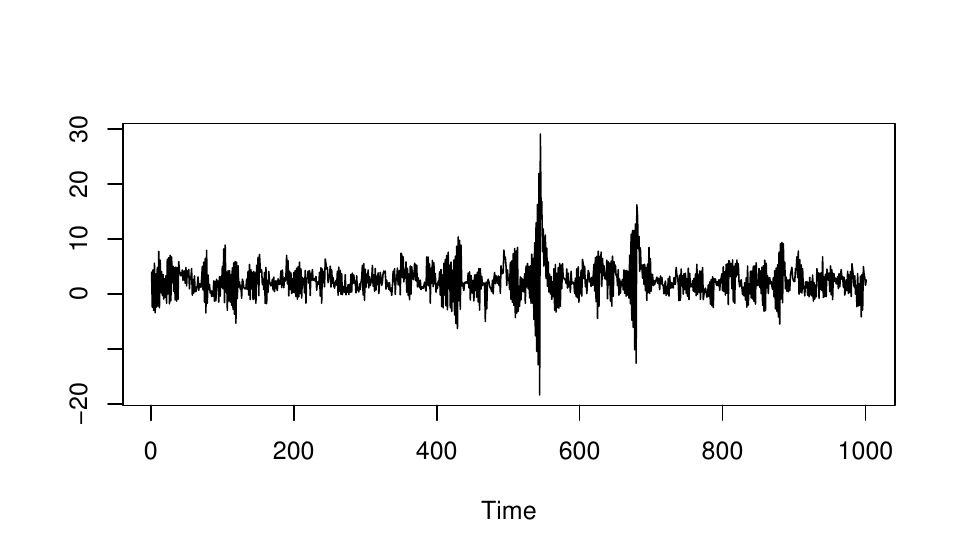}
                \caption{\footnotesize{$\left(1-\phi_{1}L\right) \left(1+\varphi_{1}L^{-1}\right) y_{t}=\varepsilon_{t}$}}
                \label{fig:C_zero_NC_pi}
\end{subfigure} \hfill 
\begin{subfigure}[b]{0.475\textwidth}
                \includegraphics[width=\textwidth]{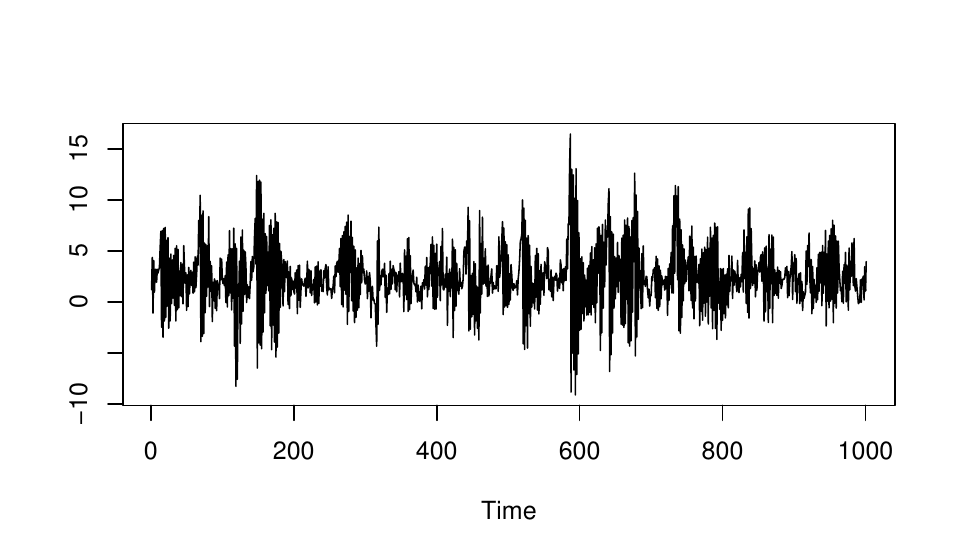}
                \caption{\footnotesize{$\left(1+\phi_{1} L\right)\left(1-\varphi_{1} L^{-1}\right)y_{t}=\varepsilon _{t}$}}
                \label{fig:C_pi_NC_zero}
        \end{subfigure}
\vskip\baselineskip
\begin{subfigure}[b]{0.475\textwidth}
                \includegraphics[width=\textwidth]{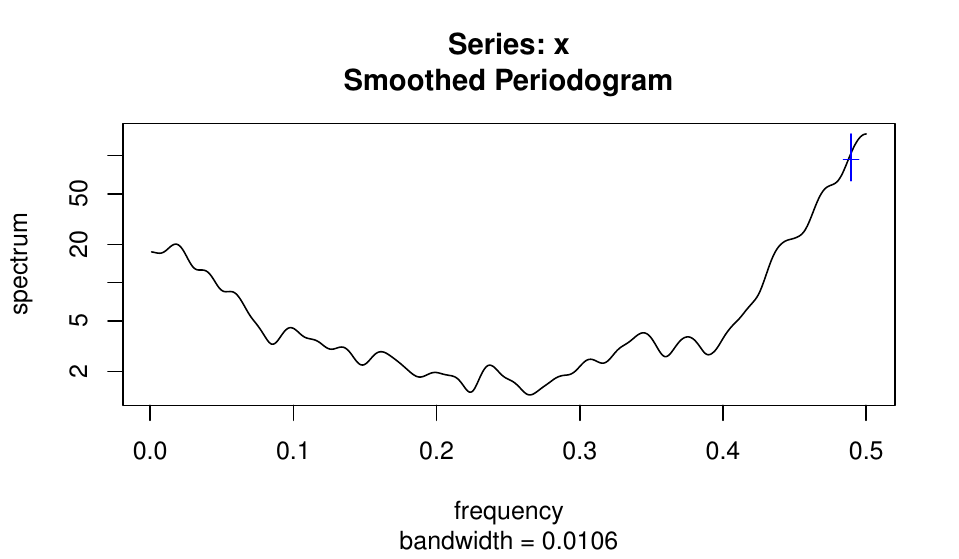}
                %\caption{\footnotesize{}}
                \label{fig:C_zero_NC_pi_perio}
        \end{subfigure}
\quad 
\begin{subfigure}[b]{0.475\textwidth}
                \includegraphics[width=\textwidth]{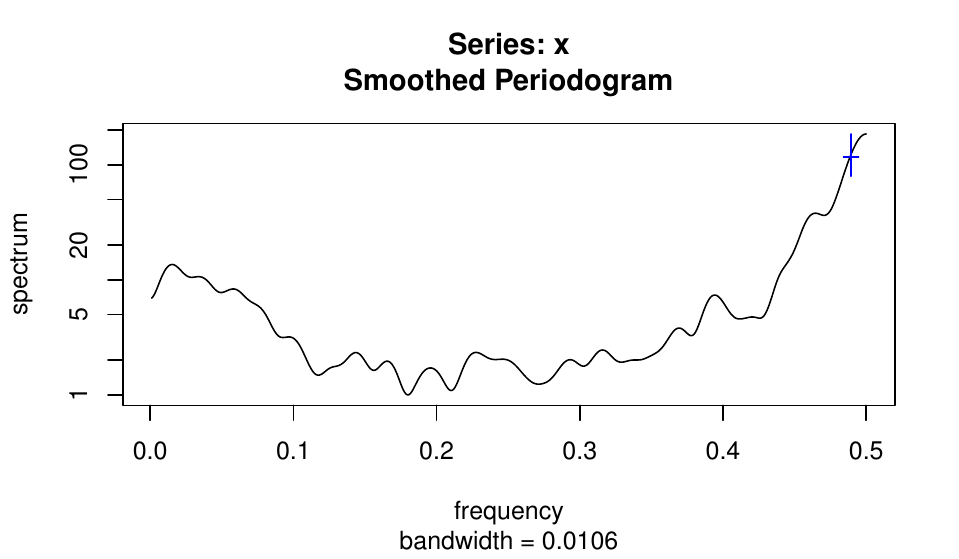}
                %\caption{\footnotesize{}}
                \label{fig:C_pi_NC_zero_perio}
        \end{subfigure}
\vskip\baselineskip
\begin{subfigure}[b]{0.475\textwidth}
                \includegraphics[width=\textwidth]{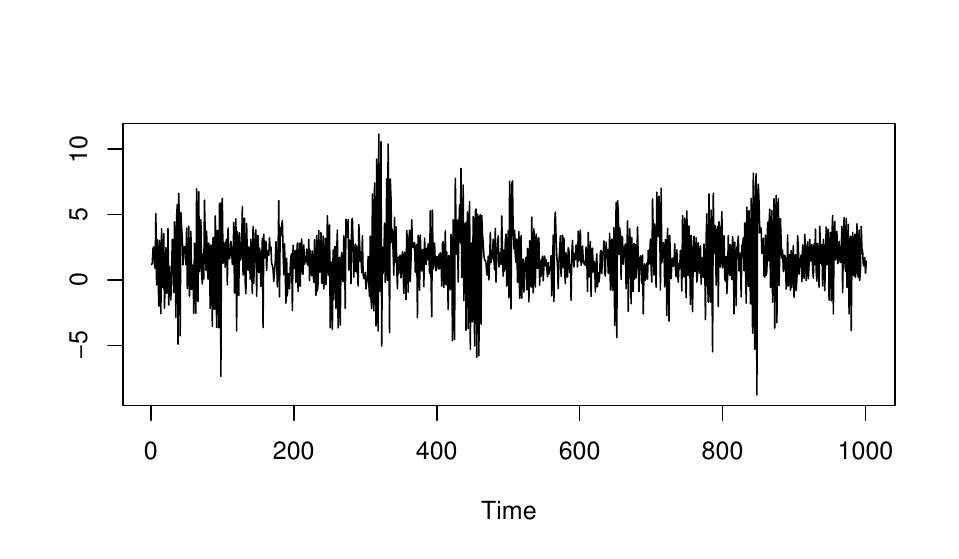}
                \caption{\footnotesize{$\left(1-\phi_{1}L\right) \left(1-2\cos\left(\frac{2\pi}{3}\right)\varphi_{1}L^{-1}+\varphi_{1}^{2}L^{-2}\right) y_{t}=\varepsilon _{t}$}}
                \label{fig:C_zero_NC_conjugate}
        \end{subfigure}
\quad 
\begin{subfigure}[b]{0.475\textwidth}
                \includegraphics[width=\textwidth]{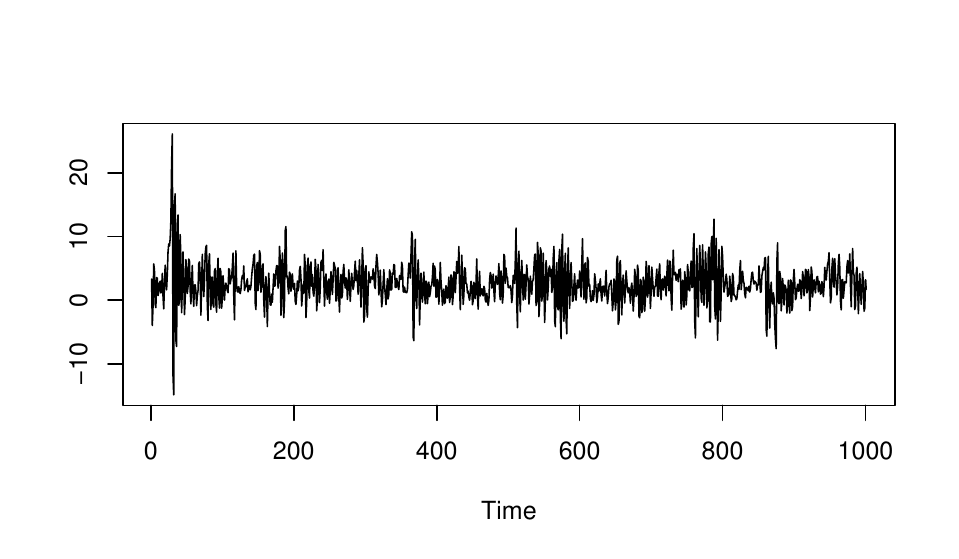}
                \caption{\footnotesize{$\left(1-2\cos\left(\frac{\pi}{2}\right)\phi_{1}L+\phi_{1}^{2}L\right) \left(1-\varphi_{1}L^{-1}\right) 
							   y_{t}=\varepsilon _{t}$}}
                \label{fig:C_conjugate_NC_zero}
        \end{subfigure}
\vskip\baselineskip
\begin{subfigure}[b]{0.475\textwidth}
                \includegraphics[width=\textwidth]{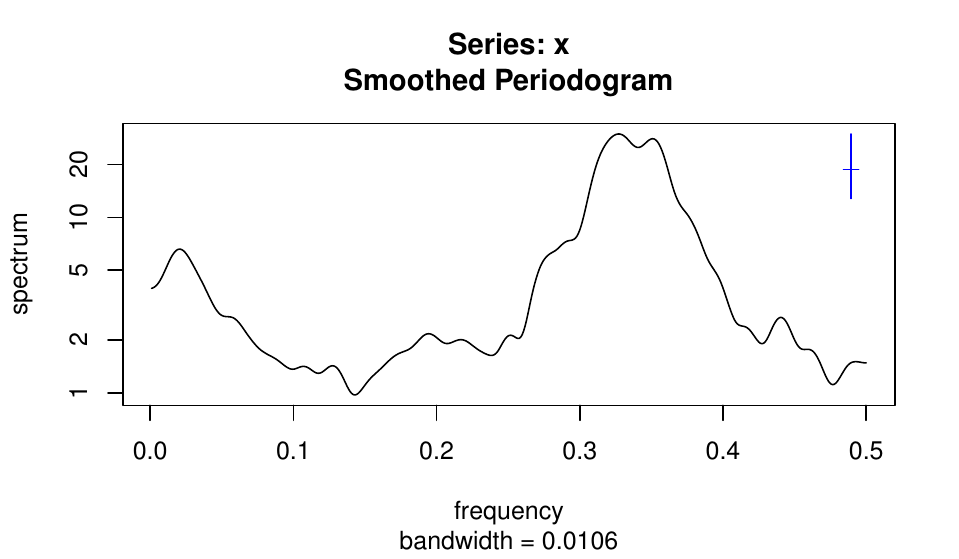}
                %\caption{\footnotesize{}}
                \label{fig:C_zero_NC_conjugate_periodogram}
        \end{subfigure}
\quad 
\begin{subfigure}[b]{0.475\textwidth}
                \includegraphics[width=\textwidth]{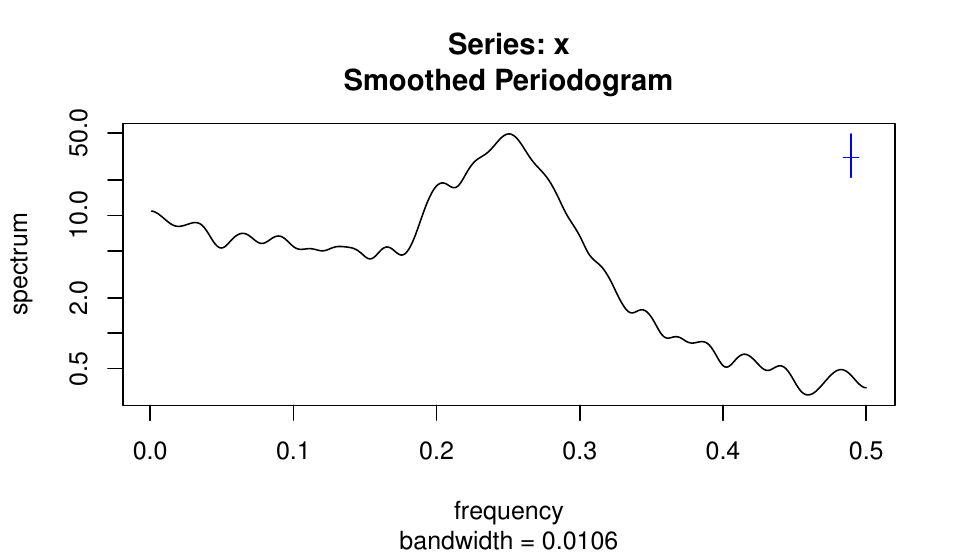}
                %\caption{\footnotesize{}}
                \label{fig:C_conjugate_NC_zero_periodogram}
        \end{subfigure}
\end{adjustbox}
\caption{Simulated MAR processes and their corresponding periodograms}
\label{fig:sim_processes_periodogram}
\end{figure}

Figure \ref{fig:sim_processes_periodogram} shows simulated data and corresponding 
periodograms of four different MAR models:
\begin{enumerate}
\item[(a)] MAR(1,1) with causal root at zero frequency and noncausal root at Nyquist frequency;
\item[(b)] MAR(1,1) with causal root at Nyquist frequency and noncausal root at zero frequency;
\item[(c)] MAR(1,2) with causal root at zero frequency and complex noncausal root (conjugate pair);
\item[(d)] MAR(2,1) with complex causal root (conjugate pair) and noncausal root at zero frequency.
\end{enumerate} 

\end{document}